\documentclass[twocolumn]{biophys-new}
\usepackage[utf8]{inputenc}
\usepackage{graphicx}
\usepackage{dsfont}
\usepackage[draft]{changes}
\usepackage[colorlinks,allcolors=cyan!70!black]{hyperref}
\usepackage{cuted}
\usepackage{xfrac}
\usepackage{changes}
\usepackage{comment}
\usepackage{widetext}
\usepackage{flushend}
\usepackage[capitalise]{cleveref}
\crefdefaultlabelformat{#2#1#3}
\usepackage{array}
\usepackage{float}
\usepackage{algorithm}
\usepackage{algpseudocode}
\usepackage{titleref}
\usepackage{wrapfig}
\usepackage{cuted}
\usepackage{subcaption}

\def\eg{\textit{e.g.}}

\def\dd{\textrm{d}}

\title{Stochastic dynamics and ribosome-RNAP interactions
in Transcription-Translation Coupling} 

\runningtitle{Stochastic ribosome-RNAP coupling model} 

\author[1]{Xiangting Li}
\author[1,2,*]{Tom Chou}
\runningauthor{Li and Chou} 

\affil[1]{Department of Computational Medicine, University of California, Los Angeles, CA 90024}
\affil[2]{Department of Mathematics, University of California, Los Angeles, CA 90024}

\corrauthor[*]{tomchou@ucla.edu}

\papertype{Article}
\begin{document}

\begin{frontmatter}
\begin{abstract}
Under certain cellular conditions, transcription and mRNA translation
in prokaryotes appear to be ``coupled,'' in which the formation of
mRNA transcript and production of its associated protein are
temporally correlated. Such transcription-translation coupling (TTC)
has been evoked as a mechanism that speeds up the overall process,
provides protection during the transcription, and/or regulates the
timing of transcript and protein formation. What molecular mechanisms
underlie ribosome-RNAP coupling and how they can perform these
functions have not been explicitly modeled. We develop and analyze a
continuous-time stochastic model that incorporates ribosome and RNAP
elongation rates, initiation and termination rates, RNAP pausing, and
direct ribosome and RNAP interactions (exclusion and binding).  Our
model predicts how \textit{distributions} of delay times depend on
these molecular features of transcription and translation. We also
propose additional measures for TTC: a direct ribosome-RNAP binding
probability and the fraction of time the translation-transcription
process is ``protected'' from attack by transcription-terminating
proteins.  These metrics quantify different aspects of TTC and
differentially depend on parameters of known molecular processes. We
use our metrics to reveal how and when our model can exhibit either
acceleration or deceleration of transcription, as well as protection
from termination. Our detailed mechanistic model provides a basis for
designing new experimental assays that can better elucidate the
mechanisms of TTC.
\end{abstract}

\begin{sigstatement}
Transcription-translation coupling (TTC) in prokaryotes is thought to
control the timing of protein production relative to transcript
formation. The marker for such coupling has typically been the
measured time delay between the first completion of transcript and
protein.  We formulate a stochastic model for ribosome and RNAP
elongation that also includes RNAP pausing and ribosome-RNAP
binding. The model is able to predict how these processes control the
distribution of delay times and the level of protection against
premature termination. We find relative speed conditions under which
ribosome-RNAP interactions can accelerate or decelerate transcription.
Our analysis provides insight on the viability of potential TTC
mechanisms under different conditions and suggests measurements that
may be potentially informative.
\end{sigstatement}
\end{frontmatter}

\section*{Introduction}

In prokaryotic cells, transcription and translation of the same genes
are sometimes ``coupled'' in that the first mRNA transcript is
detected coincidentally with the first protein associated with that
transcript. This observation suggests proximity of and interactions
between the ribosome and the RNA polymerase (RNAP).  Ribosome-RNAP
interactions in prokaryotes are thought to maintain the processivity
of RNA polymerase (RNAP) by physically pushing it out of the
paused, backtracking state \cite{Stevenson-Jones2020}. Higher processivity can
also suppress cleavage and error correction of the mRNA transcript,
inducing the RNAP to incorporate nucleotides and continue
transcription.  Transcription-translation coupling (TTC) may also play
an important role in protecting mRNA from premature transcription
termination \cite{Chalissery2011,Lawson2018,Kohler2017apr}.  This
protection might arise from steric shielding of the elongation complex
by the leading ribosome, preventing attack by Rho
\cite{Kohler2017apr,Ma2015feb}.

Evidence for TTC has come from two types of experiments.  The first
is ``time-of-flight'' experiments that quantify the time delay
between first detection of a complete transcript and a complete
protein.  For example, IPTG-induced LacZ completion experiments
measure the mean time of mRNA completion $\overline{T}_{\rm RNAP}$ and
the mean time of protein completion by the leading ribosome
$\overline{T}_{\rm rib}$, with the latter measured from the time of
first RNAP engagement \cite{Proshkin2010,Iyer2018,Vogel1994}.  Since
the transcript length $L$ is known, the effective velocities of the
RNAP and ribosome over the entire transcript can be estimated by
\begin{equation}
    \overline{V}_{\rm RNAP} = \frac{L}{\overline{T}_{\rm RNAP}},\,\,\,  
\overline{V}_{\rm rib} =\frac{L}{\overline{T}_{\rm rib}}.
  \label{eq:effective_velocity}
\end{equation}
These measurements are performed at the population level, averaging
the time-dependent signal from many newly formed transcripts and
corresponding proteins. Thus, the individual molecular coupling
mechanisms between RNAP and ribosomes cannot be resolved by the time
delay unless single molecule time-of-flight experiments can be
designed.

Another class of experiments uses a variety of \textit{in vitro} and
\textit{in vivo} assays to probe direct and indirect molecular
interactions between RNAPs and ribosomes
\cite{Fan2017,Kohler2017apr,Mooney2009,Saxena2018}. However, while
providing context for what types of molecular interactions are
possible, these experiments have not unequivocally observed direct
binding \textit{in vivo}.

Two modes of interaction between the leading ribosome and the RNAP
have been proposed. One mode of interaction is through a ``collided
expressome'' in which the ribosome and RNAP are held in close
proximity \cite{Fan2017,Kohler2017apr} by direct association.  The
second coupling mode occurs through a larger complex in which
ribosome-RNAP interactions are mediated by the protein NusG
\cite{Mooney2009,Saxena2018}.  There have been no reports that this
mode alters elongation velocities or RNAP processivity but it has been
shown that the NusG-coupled expressome can inhibit Rho-induced
premature transcription termination \cite{Burmann2010}.
\begin{figure}[!tbh]
    \centering
    \includegraphics[width=3.25in]{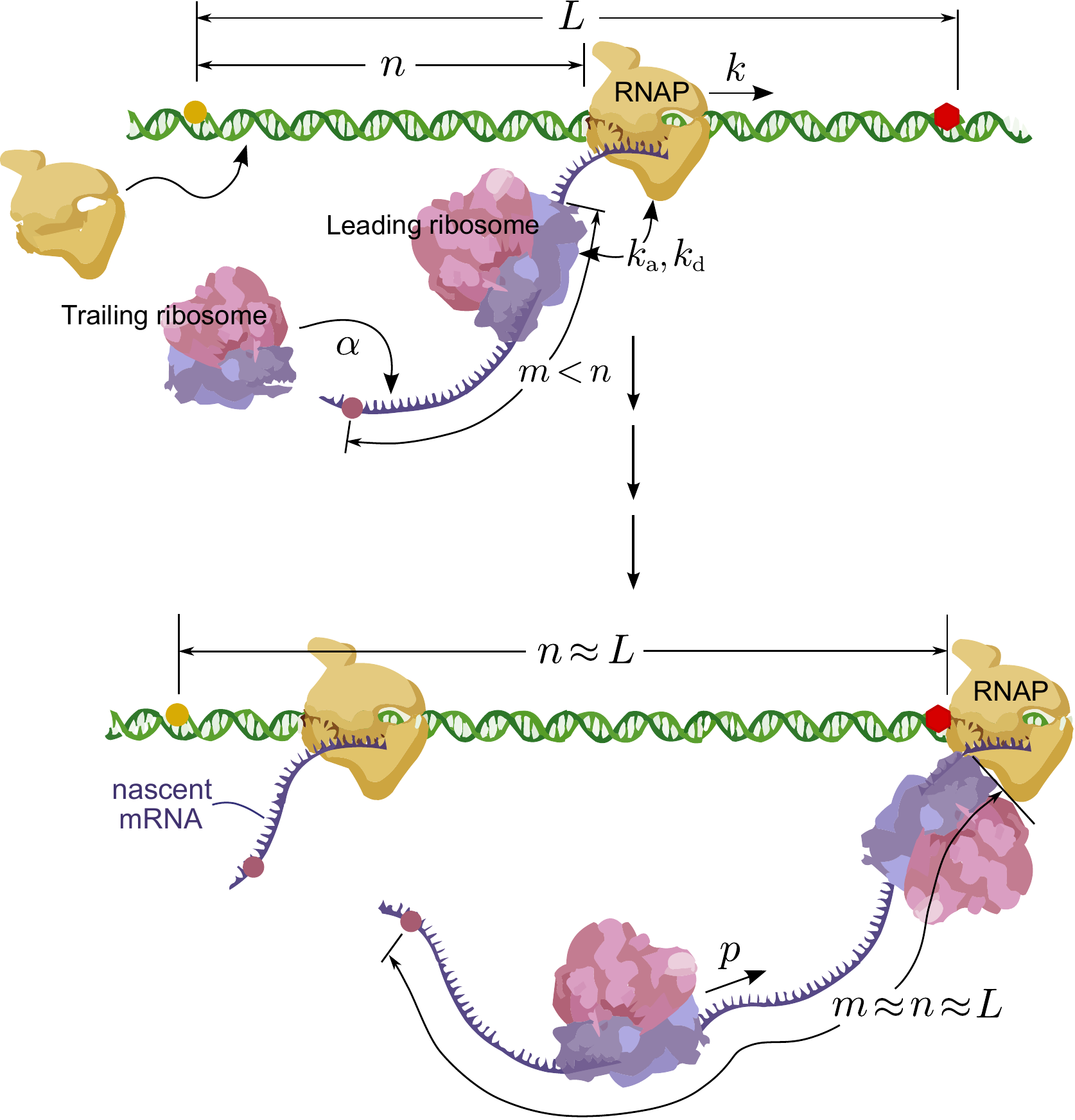}
    \caption{Schematic of translation of a nascent mRNA transcript
      (polypeptide not shown). Ribosomes translocate at rate $p$ while
      RNAPs elongate at rate $q$.  Top: The transcript associated with
      the RNAP at position $n$ along the gene is shown with a leading
      ribosome at position $m$ along the mRNA. Ribosomes attach to
      open initiation sites at rate $\alpha$. Bottom: A nearly
      complete transcript is shown. If the leading ribosome has caught
      up to the RNAP and $m$ is close to $n$, the two may bind with
      rate $k_{\rm a}$ to form a ``collided expressome.''
      Ribosome-RNAP complexes can spontaneously dissociate with rate
      $k_{\rm d}$.  We assume the leading ribosome ``terminates'' upon
      reaching the stop codon (not shown). Protein-mediated
      expressomes (not shown) form larger complexes that can
      accommodate longer mRNA segments within it.}
\label{fig:introduction}
\end{figure}

Both potential coupling mechanisms require at least some moments of
close proximity between the RNAP and the leading ribosome during the
simultaneous transcription-translation process (see
Fig.~\ref{fig:introduction}), followed by recruitment of NusG for the
NusG-coupled expressome mechanism.
%
%
The ribosome-RNAP proximity requirement can be met if the ribosome
elongation speed is, on average, faster than that of the RNAP.  Even
if the ribosome is fast, proximity also depends on initial condition
(ribosome initiation delay after RNAP initiation) and the length of
the transcript $L$. Moreover, both RNAPs and ribosomes are known to
experience, respectively, pausing through backtracking \cite{Zuo2022}
and through ``slow codons'' for which the associated tRNA is scarce
\cite{Lakatos2004}.

%

A number of open questions remain. In the ``strong coupling'' picture,
the ribosome and RNAP are nearly always in contact and the speed of
the ribosome is thought to limit that of the RNAP. Since under typical
growth conditions, a ribosome translocates at the same 45nt/s speed as
RNAP, the strong coupling picture provides an attractive explanation
for the slowdown from 90nt/s in rRNA transcription to 45nt/s in mRNA
transcription \cite{Vogel1994}.  Administration of antibiotics to slow
down translation also slowed down transcription.

However, other experiments have shown that the distance between the
ribosome and RNAP can be large most of the time, leading to a ``weak
coupling'' picture \cite{Chen2018,Zhu2019}. The biological role of
weak coupling is unclear since any shielding provided by the ribosome
would be limited and ribosome and RNAP speeds could be independently
modulated.  Even though direct ribosome-RNAP interactions may still
arise after an RNAP has stalled for a sufficiently long time, any
apparent ribosome-RNAP coordination would be largely coincidental.

Besides the strong and weak coupling dichotomy, another unknown is
whether there are direct molecular interactions between the leading
ribosome and RNAP. Although experiments to probe such interactions
during normal transcription and translation \textit{in vivo} will be
difficult to design, our model can provide easier-to-measure
indicators of molecular coupling.

\begin{figure*}[h!]
    \centering
    \includegraphics[width=5.6in]{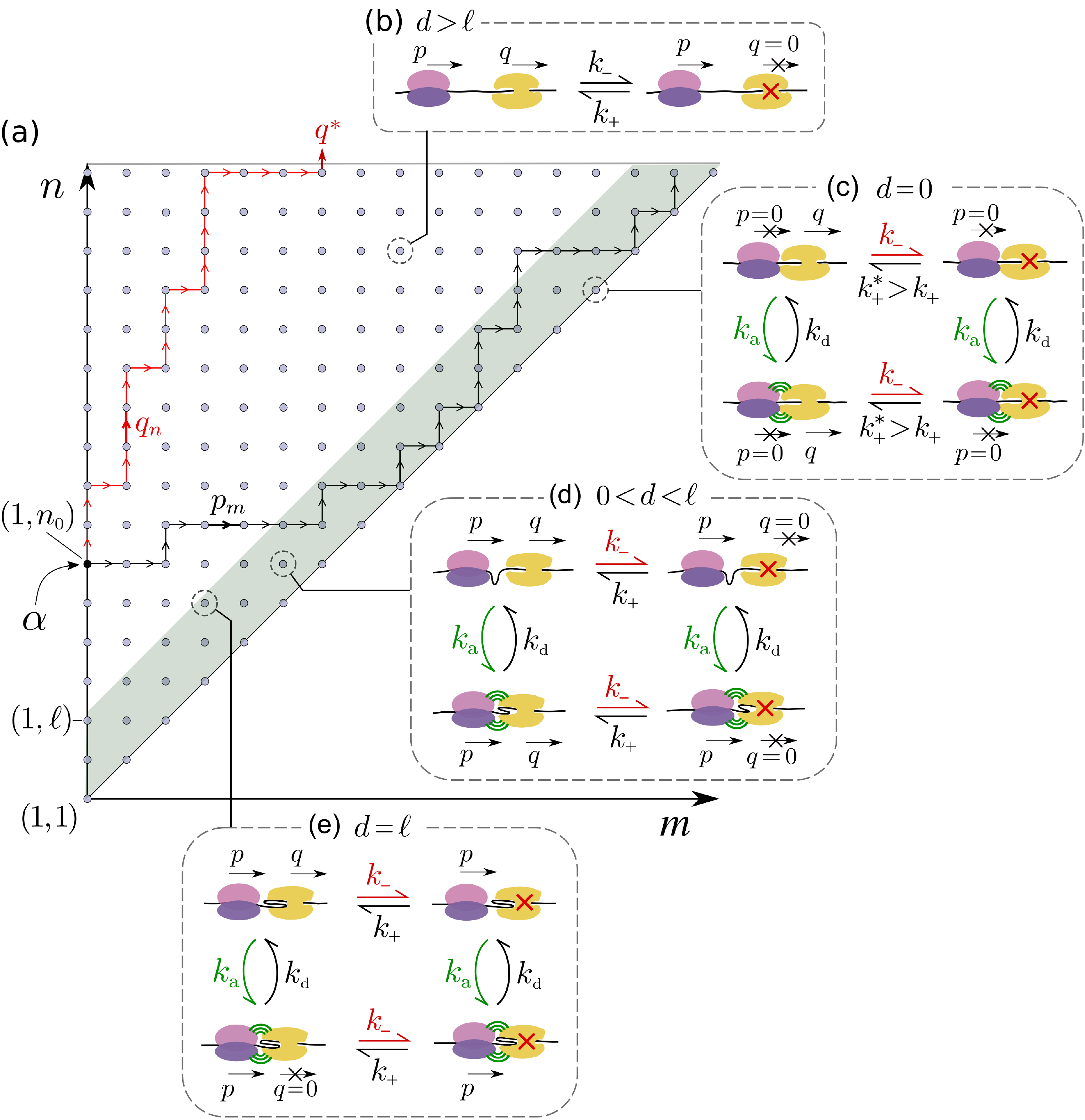}
    \caption{(A) State space of the stochastic model defined in terms
      of the leading ribosome and RNAP positions $(m,n)$. The initial
      time $t=0$ is defined as the time RNAP first produces a ribosome
      initiation site, starting the system in $(m=0, n=1)$. For $t>0$,
      as the RNAP is elongating, the first ribosome binds at rate
      $\alpha$. Here, a ribosome binds after the RNAP first reaches
      position $n=n_{0}$. Red and blue trajectories indicate scenarios
      in which the RNAP is relatively fast and slow,
      respectively. Within each position $(m,n)$ exist internal
      molecular microstates states. (B) In the ``interior'' states
      $n-m > \ell$ ($\ell = 2$ in this example), the ribosome and RNAP
      are too distant to be bound, and only stalled and processing
      RNAP states arise, with transition rates $k_{\pm}$ between
      them. (C) When $n=m=0 \leq \ell$, the ribosome and RNAP are
      adjacent without any intervening mRNA, allowing them to
      associate with rate $k_{\rm a}$. The RNAP can be in either
      stalled or processive states. In the stalled state, whether
      associated or not, the adjacent volume-excluding ribosome
      entropically ``pushes'' the stalled RNAP, catalyzing its
      transition to a processive state so that $k_{+}^{*} > k_{+}$.
      (D) When $0<n-m < \ell$, the ribosome and the RNAP are close
      enough to bind with rate $k_{\rm a}$. Here, the intervening mRNA
      dissipates the entropic pushing (so that the stalled RNAP $\to$
      processing RNAP transition rate is $k_{+}$) and also allows an
      RNAP in the processive state to elongate with rate $q$,
      regardless of whether it is bound to the ribosome. (E) Only when
      the ribosome and the RNAP are separated by $d=\ell$ is a bound
      RNAP prevented from processing as this would reel in more mRNA
      than can be fit inside a collided expressome. Molecular binding
      prevents complexed ribosome and RNAP to be separated by more
      than $\ell$ mRNA codons.}
\label{fig:mechanism}
\end{figure*}

To help resolve the puzzles discussed above, provide a quantitative
way to explore different molecular mechanisms that may contribute to
TTC, and generate predictions that can be compared to experimental
observations, we formulate a stochastic model that combines a number
of known molecular mechanisms from transcription, keeping track of
ribosome and RNAP states and positions along the gene.  While an
earlier model combined transcription and translation in prokaryotes
\cite{Makela2011}, it did not explicitly incorporate mechanisms of
direct transcription and translation coupling and only assumed simple
volume exclusion between the RNAP and the leading ribosome.

Here, we explicitly allow for RNAP pausing and direct association and
dissociation of the ribosome-RNAP complex.  The typical assay used to
probe TTC involves measuring the time delay $\Delta T = T_{\rm
  rib}-T_{\rm RNAP}$ between the completion of mRNA and its associated
protein.  Although time delays can be used as a metric for defining
transcription-translation coupling, absence of delay is a necessary
but not sufficient condition for direct ribosome-RNAP coupling.  A
small mean delay time can arise simply from coincidental proximity of
the ribosome to the RNAP at the time of termination. On the other
hand, a significant time delay may indicate an uncoupled process
especially if the delay is variable and cannot be controlled
\cite{Johnson2020}.  After formulating our model, we construct
additional metrics that better define TTC. However, since the time
delay is the most experimentally measurable quantity, we will still
derive and compute the full probability density of delay times
$\rho(\Delta T)$.


\section*{Model and Methods}
\begin{table*}[h!]
\caption{Model parameters}
\label{tab:paras}
\centering
\begin{threeparttable}
\begin{tabular}{c l >{\centering\arraybackslash}m{6cm} l}
\hline
Params. & Description & Typical values\tnote{a} & Refs. \\\hline
$\alpha$ & translation initiation rate & $\sim 0.01-10.0$ $s^{-1}$ & \cite{Dai2016dec,Johnson2020,Shaham2017nov,Kennell1977jul} \tnote{b}  \\
$L$ & gene and transcript length & $L\in  \mathds{Z}^{+},\, L\sim 300$ & \cite{Xu2006}  \\
$m$ & ribosome position from mRNA 5'& $m\in \mathds{Z}_{\geq 0},\, 0\leq m\leq L$  & --  \\
$n$ & RNAP position from mRNA 5' & $n\in \mathds{Z}_{+},\, m \leq n\leq L$  & --  \\
$p$ & free ribosome translocation rate & $\sim 15$ codons/s & \cite{Johnson2020,Young1976,Proshkin2010,Zhu2016}  \\
$q$ & free processing RNAP transcription rate & $\sim 30$ codons/s & \cite{Proshkin2010,Iyer2018,Vogel1994,Epshtein2003may}\tnote{c}  \\
$k_{-}$ & processive RNAP $\to$ paused RNAP rate & $\sim 0.4$ $s^{-1}$ & \cite{Neuman2003} \tnote{d} \\
$k_{+}$ & paused RNAP $\to$ processive RNAP rate  & $\sim 0.3$ $s^{-1}$ & \cite{Neuman2003}  \\
$k_{+}^*$ & paused RNAP $\to$ processive RNAP rate (pushed) & $k_{+}^*=k_{+}\exp(E_+), E_{+}\geq 0$    & estimated  \\
$k_{\rm a},k_{\rm d}$ & ribosome-RNAP association, dissociation rates & 
$k_{\rm d}=k_{\rm a}e^{-E_{\rm a}}$, $E_{\rm a} \sim 3-7$ & \cite{Fan2017}  \\
$\ell$ & maximum mRNA length in bound complex & $\sim 4-6$ codons  & \cite{Wang2020aug}\tnote{e}  \\
\hline
\end{tabular}
\begin{tablenotes}
\item[a] Ribosome and RNAP positions are measured in numbers of
  nucleotide triplets (codons) from the 5' end of the nascent mRNA.
  For simplicity, we assume RNAP and ribosome initiation sites are
  coincident along the sequence.
\item[b] The translation initiation rate $\alpha$ depends on ribosome
  availability and varies significantly across the genome
  \cite{Shaham2017nov,Siwiak2013sep}.  The median transcription
  initiation time is estimated to be $15-30$ seconds. For LacZ
  induction methods used in experiments, the initiation rates were
  assumed to be quite high. The definition of starting time depends on
  the experimental protocol and measurement.  In \cite{Johnson2020},
  the initiation time was neglected. In \cite{Dai2016dec}, the total
  time for initiation steps--including IPTG penetration, LacI
  depression, transcription initiation, and translation
  initiation--was measured to be around 10 seconds.  Slow translation
  initiation can be compensated for by transcription arrest near the
  5' proximal region of the gene \cite{Hatoum2008}, allowing for a
  smaller $n_{0}$ (see Fig.~\ref{fig:mechanism}). In our simulations,
  we set $\alpha = 1$/s.
\item[c] Typical noninteracting RNAP transcription rates are
  $\bar{q}\sim 15$ codons/s. Since typically $k_{+}/(k_{+}+k_{-})\sim
  1/2$, we use typical values $q \sim 30$ codons/s for the unimpeded
  transcription rate of processing RNAP.
\item[d] The pausing probability along an RNAP trajectory has been
  measured as $\sim 0.87$ per 100 nucleotides. By using the estimated
  mean RNAP velocity of $~\sim 15$ codons/s, we convert this
  probability to a pausing rate $k_{-}\approx 0.4$/s.
\item[e] The typical interaction range $\ell$ will be approximated by
  the maximum stored length of mRNA in a complex.  For collided
  expressomes, $\ell\sim 4$ codons, while for NusG-mediated complexes,
  $\ell \sim 8$ codons since its larger size can accommodate more
  intervening mRNA.
\end{tablenotes}
\end{threeparttable}
\end{table*}

Based on existing structural and interaction information, we formulate
a continuous-time Markov chain to model ribosome and RNAP kinetics. As
shown in Fig \ref{fig:introduction}, we describe the position of the
head of the leading ribosome along the nascent transcript by $m
=0,1,\ldots, L$, where $0$ denotes a ribosome-free transcript.  We
also track the length of the nascent mRNA transcript that has cleared
the exit channel of the RNAP through the discrete variable $n = 1,2,
\ldots, L$.  The positions are described in terms of triplets of
nucleotides corresponding to codons, the fundamental step size during
ribosome elongation.  Here, $L$ is the length of the gene, typically
about $L\sim 300$ codons.  We carefully choose the definition of $m$
and $n$ so that ribosome and RNAP sizes are irrelevant and that the
difference $d\equiv n-m$ precisely describes the length of the free
intervening mRNA between them. While this assumption is certainly not
true due to shorter leading and termination segments specific to
translation, the slight differences in length are assumed negligible,
or are subsumed in effective translation initiation rates.  Therefore,
$0\leq m \leq L$ and $1\leq n\leq L$, where $n=L$ is interpreted as a
completion mRNA and $m=L$ is interpreted as a completed polypeptide.
This triangular state-space structure has arisen in related stochastic
models of interacting coordinates in one-dimension
\cite{Chou2007,Zuo2022,Kolomeisky2022}.

Here, within each positional state $(m,n)$, the leading ribosome and
RNAP can exist in different internal configurations describing their
molecular states.  The RNAP at site $n$ can switch between two states,
a processive state and a paused state.  In the processive state, the
RNAP can move forward by one codon at rate $q_{n}$ or it can
transition to a paused or ``backtracking'' state with stalling rate
$k_{-}$.  The RNAP elongation rate can also depend on its position $n$
through different abundances of corresponding nucleotides. For
simplicity, we assume that RNAPs in the backtracking state are fixed
and do not elongate ($q_{n}=0$) but may transition back to the
processive state with ``unstalling'' rate $k_{+}$.  The waiting time
distributions in the processive and paused states are exponential with
mean $1/k_{-}$ and $1/k_{+}$, respectively. The leading ribosome at
site $m$ will be assumed to always be in a processive state with
forward hopping rate $p_{m}$ if and only if the next site $m+1$ is
empty (not occupied by the downstream RNAP). In general, the ribosome
translation rate can depend on the position $m$ through the codon
usage at that site.

%
%

When the distance between the leading ribosome and the RNAP is within
an interaction range $\ell$, ($d\equiv n-m\leq \ell$), they may bind
with rate $k_{\rm a}$ to form a collided expressome and dissociate
with rate $k_{\rm d}$ (Eq.  \ref{eqn:coupling} and
\ref{eqn:uncoupling}). To enumerate internal states that are
associated/disassociated and processing/backtracking, we define
$(a,b)\in \left\{ 0,1 \right\}^2$ such that $a=1$ refers to an
associated, or ``bound'' ribosome-RNAP complex, and $b=1$ refers to an
RNAP in a backtracking, or a ``paused'' or ``stalled'' state. When
$a=0$, the ribosome is not bound to the RNAP, and when $b=0$, the RNAP
is in the processive state.  The state space of our discrete
stochastic model is given by $\left\{ (m,n,a,b): 1 \leq m \leq n \leq
L, a\in \{ 0,1 \}, b\in \{0,1\} \right\}$, with \, \\
$\left\{(0,n,0,b): 1 \leq n \leq L, b=\pm 1 \right\}$ representing
ribosome-free configurations.

Other than steric exclusion (which constrains $m\leq n$) and
ribosome-RNAP association and dissociation, we incorporate a
contact-based RNAP ``pushing'' mechanism.
%
%
The processing ribosome can directly push (powerstroke) against a
stalled RNAP and/or reduce the entropy of a backtracking RNAP to bias
it towards a processive state. A similar mechanism arises in RNAP-RNAP
interactions as discussed in \cite{Zuo2022}. To quantify this pushing
mechanism, we simply modify the paused-to-processive RNAP ($b=1 \to
b=0$) transition rate from $k_{+}$ to $k_{+}^{*} \equiv k_{+}e^{E_{+}}
> k_{+}$ whenever the ribosome abuts the RNAP ($d\equiv n-m = 0$). The
enhanced rate arises from a reduction $E_{+}$ in the total transition
free energy barrier provided by the adjacent ribosome. Typical model
parameters relevant to prokaryotic transcription and translation are
listed in Table \ref{tab:paras}.




The length $\ell$ may influence direct molecular
coupling and stochastic dynamics of transcription.  \textit{In
  vitro} studies of ribosome and RNAP structure provide constraints on
the configuration space accessible to coupled expressomes.  Wang et
al. \cite{Wang2020aug} found that collided expressomes are stable only
when the spacer mRNA between the ribosome and the RNAP is $\sim~12-24$
nucleotides ($\sim 4-8$ codons).  Because the intervening mRNA must be
at least $12$ nucleotides to extend beyond the RNA exit channel of the
RNA polymerase, the free intervening RNA within an intact collided
expressome can vary between 0 and 12 nucleotides.  In contrast, the
NusG-mediated expressome can accommodate $\sim 24-30$ free mRNA
nucleotides. RNA looping might allow for even longer spacer mRNA, but
there has so far been no \emph{in vivo} evidence that collided
expressomes exist with mRNA loops.

Since mRNA is flexible, we can also assume that $k_{\rm d}$ is
constant for $d\equiv n-m \leq \ell$.  The association rate $k_{\rm
  a}$ may be dependent on the distance $d=n-m$ between the ribosome
and the RNAP; for example, a distance-dependent association rate might
take the form $k_{\rm a}(n-m) \approx k_{\rm a}(\ell) [(\ell
  +\xi)/(n-m+\xi)]^3$, where $[(n-m)+\xi]^{-3}$ represents the
effective volume fraction of the leading ribosome and $\xi$ is the
configuration flexibility of ribosome-RNAP binding when they are
close. If we adopt such a distance-dependent $k_{\rm a}$, we would
also have to let the ratio $p/q$ be dependent on $(n-m)$ in order
conserve free energy during approach and binding steps.  To simplify
matters, we will assume $\xi \gg \ell$ and take $k_{\rm a}$ to be a
constant for $d\equiv n-m \leq \ell$ and zero for $d\equiv n-m >
\ell$.

The overall kinetics of the internal states pictured in the insets of
Fig.~\ref{fig:mechanism} can be explicitly summarized by considering the 
intervening mRNA length $d=n-m$ between RNAP
and the ribosome.
\begin{figure}[!tbh]%
  \centering
  \includegraphics[width=3.2in]{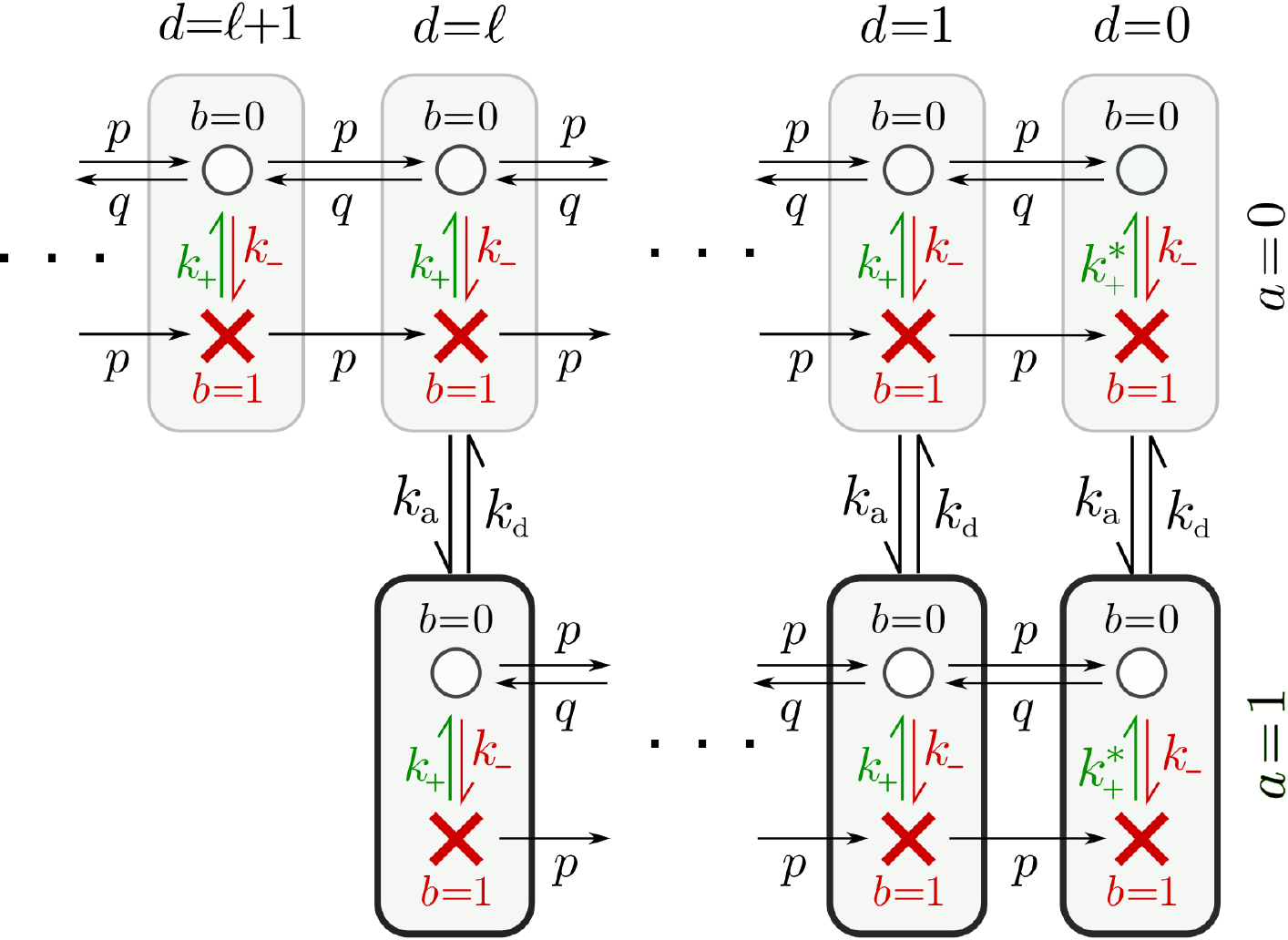}
  \caption{Internal state space associated with different values of
    the ribosome-RNAP distance $d=n-m$.  Open circles ($b=0$)
    represent states with processing RNAP that allow $d$ to increase
    and decrease with rate $q$ and $p$, respectively.  Processing
    states can transition with rate $k_{-}$ to states that contain a
    paused RNAP (red crosses, $b=1$, $q=0$).  When $0\leq d\leq \ell$,
    transitions between coupled and uncoupled states occur at rate
    $k_{\rm d}, k_{\rm a}$.  When $d=\ell$ and $a=1$, the coupled RNAP
    cannot proceed, even if it is processive, without first
    dissociating from the ribosome.  Only after detachment ($a=0$) can
    the separation exceed $\ell$, during which binding cannot occur
    and the ribosome and the RNAP process independently without the
    tethering constraint. Thus, bound states with $a=1$ (and $0\leq
    d\leq \ell$) describe a kinetic trap in which the ribosome and
    RNAP are tethered by the $\ell$-length mRNA and overall
    transcription can occur only through inchworming.  Whether bound
    or unbound, when $d=0$, the ribosome increases the RNAP unstalling
    rate from $k_{+}$ to $k_{+}^{*} = k_{+}e^{E_{+}}$.}
  \label{fig:distance_state_space}
\end{figure}
Fig.~\ref{fig:distance_state_space} explicitly depicts the transitions
as a function of $d$.  Since in our model, the maximum length of mRNA
that can fit within the complex is $\ell$, a processing ribosome-bound
RNAP at $n=m+\ell$ cannot advance to lengthen the already compressed
transcript. The only way a coupled state $a=1$ with $d=\ell$ can reach
any state where $d > \ell$ is for the ribosome and RNAP to first
dissociate (we assume dissociation rates in all $d=n-m$ states remain
constant at $k_{\rm d}$). Molecular coupling effectively slows down
transcription by preventing RNAP elongation in the $a=1,d=\ell$ state.
Such ribosome-mediated slowing down of transcription has been proposed
in previous studies \cite{Stevenson-Jones2020,Kohler2017apr}.

We now list all allowed transitions in the $\omega \coloneqq \{m,n,a,b\}$
state space of our continuous-time stochastic Markov model.  The
probability that an allowable transition from state $s$ to state $s'$
occurs in time increment $\dd t$ is $r(\omega' \vert \omega)\dd t$ where the
complete set of rates is given by
\begin{align}
  r(1,n,a,b\,\vert\, 0,n,a,b) & = \alpha, && 1 \leq n \leq L, \label{eqn:initiation} \\
  r(m+1,n,a,b\,\vert\, m,n,a,b) & = p_{m}, && 1 \leq m \leq n-1, \label{eqn:ribosome_translocate} \\
  r(m,n+1,0,0\,\vert\, m,n,0,0) &= q_{n}, && m \leq n \leq L-1, \label{eqn:RNAP_translocate0} \\
  r(m,n+1,1,0\,\vert\, m,n,1,0) & = q_{n}, && 0 \leq d \leq \ell-1,  \label{eqn:RNAP_translocate} \\
  r(m,n,1,b\,\vert\, m,n,0,b) & = k_{\textrm{a}}, && 0\leq d \leq \ell,  \label{eqn:coupling} \\
  r(m,n,0,b\,\vert\, m,n,1,b) & = k_{\textrm{d}}, && \:  \label{eqn:uncoupling} \\
  r(m,n,a,1\,\vert\, m,n,a,0) & = k_{-}, && \:  \label{eqn:pausing} \\
  r(m,n,0,0\,\vert\, m,n,0,1) & = k_{+}, && \:   \label{eqn:restarting0}\\
  r(m,n,1,0\,\vert\, m,n,1,1) & = k_{+}^{*}, && a=1, m=n.  \label{eqn:restarting}
\end{align}
Using these rules, we performed event-based stochastic simulations
\cite{Bortz1975,Gillespie1977dec} of the model as detailed in Appendix
\ref{SS} of the Supplemental Information (SI).
%
%
For completeness, the master equation associated with our model is
also formally given in Appendix \ref{numerical}.

\paragraph*{Construction of time delay distribution.}
Our model allows for explicit calculation of the distribution
$\rho(\Delta T)$ of time delay $\Delta T$.  To find $\rho(\Delta T)$,
we first find the distribution of ribosome positions $m(T_{i})$ at the
moment $T_{i}\equiv T(n=i)$ the RNAP first reaches site $i$. 
$T_{L}\equiv T_{\rm RNAP}$ denotes the instant the mRNA is completed.
%
%
The initial value $m(T_{1})=0$ is known because immediately after
initiation of RNAP at site $n=1$, the ribosome is not yet present but
is trying to bind at a rate of $\alpha$.  As detailed in
Appendix~\ref{conditional_distribution}, we can iteratively find the
distribution of $m(T_{i+1})$ given that of $m(T_{i})$.  By the same
method, the distribution of association values $a(T_{\rm RNAP})$ at
the instant of RNAP completion can be computed.  After constructing
the probability distribution $\mathbb{P}(m, a, b\,\vert \, t=T_{\rm
  RNAP})$, we can construct the probability density $\rho(\Delta T)$
of the mRNA protein time delay $\Delta T \equiv T_{\rm rib} - T_{\rm
  RNAP}$ by evaluating the distribution of times required for the
ribosome to catch up by reaching $m=L$.

Although we are able to construct the whole distribution of delay
times that might provide a more resolved metric, especially if
single-molecule assays can be developed, a short time delay is a
necessary but not sufficient condition for TTC. To provide direct
information on molecular ribosome-RNAP interactions, we construct
additional metrics.

\paragraph*{Coupling indices.}
To more explicitly quantify direct \textit{molecular} coupling, we also
define the \textit{coupling coefficient} $C$ by
\begin{equation}
  C \equiv  \mathbb{P}(a\,\vert\, t=T_{\rm RNAP}),
  \label{eqn:coeff-coupling}
\end{equation}
the probability that the ribosome is associated with the RNAP ($a=1$)
at the moment the mRNA transcript is completed. The coupling parameter
$C$ provides a more direct measure of molecular coupling and further
resolves configurations that have short or negligible delays. While
delay time distributions do not directly quantify ribosome-RNAP
contact, the coupling coefficient $C$ does not directly probe the
trajectories or history of ribosome-RNAP dynamics.

To also characterize the history of ribosome-RNAP interactions, we
quantify TTC by the fraction of time $F_{T}$ that the ribosome
``protects'' the RNAP across the entire transcription process.  There
are different ways of defining how the transcript is protected. While
both modes of TTC are proposed to shield the mRNA from premature
termination, neither has been directly observed \textit{in vivo}.  We
assume that a termination protein has size $\sim \ell_{\rm p}$ and
that if the ribosome and RNAP are closer than $\ell_{\rm p}$, the
termination factor is excluded.  Thus, we define the protected time as
the total time that $d< \ell_{\rm p}$ codons, divided by the time to
complete transcription:

\begin{align}
  F_{T} & =\frac{\left\lVert\left\{t:(n_t-m_t<\ell_{\rm p})\right\}\right\rVert}{T_{\rm RNAP}}.
\label{eqn:protected_1}
\end{align}
Since the transcription-termination protein Rho has an mRNA footprint
of about $80$nt, $\ell_{\rm p} \approx 27$ codons
\cite{Koslover2012nov}.  The protected-time fraction $F_{T}$ provides
yet another metric for TTC that measures the likelihood of completion.

Using these metrics and the effective velocities $\overline{V}_{\rm
  rib}$ and $\overline{V}_{\rm RNAP}$, we will explore the biophysical
consequences of our model.  Simple limits are immediately apparent.
If the free ribosome translocation rate is much greater than the free
RNAP transcription rate and $\alpha \gg L/\bar{q}$, the ribosome,
for much of the time, abuts against the RNAP, inducing it to
transcribe at rate $qk_{+}^{*}/(k_{+}^{*}+k_{-}) \equiv
\bar{q}^{*}$. Here, we predict an expected delay $\Delta \overline{T}
\approx 0$, $C\approx k_{\rm a}/(k_{\rm a}+k_{\rm d})$, and $F_{T}
\approx 1$.  If the ribosome is slow and $p\ll \bar{q}$, the ribosome
and RNAP are nearly always free, $\Delta\overline{T}\approx 1/\alpha +
L(1/p-1/\bar{q})$, $C\approx 0$, and $F_{T}\approx 0$.  However, when
$p$ is intermediate, more intricate behavior can arise, including
tethered elongation and transcription slowdown.  In the next section,
we focus on the intermediate translation rate regime and show how the
effective velocities defined in Eq.~\ref{eq:effective_velocity} and
$F_{T}$ characterize the functional dynamics of TTC and $C$
characterizes the intrinsic properties of TTC, even though
$\rho(\Delta T)$ remains the most easily measurable property of TTC.




\section*{Results and Discussion}

Here, we present analyses of solutions to our model obtained from
numerical recursion and Gillespie-type kinetic Monte Carlo simulations
detailed in Appendices \ref{SS}, \ref{numerical}, and
\ref{conditional_distribution} of the SI.  Predictions derived from
using different parameter sets are compared, and mechanistic
interpretations are provided.

\subsection*{Comparison of coupling indices}

\begin{figure}[!t]
    \centering
    \includegraphics[width=3.35in]{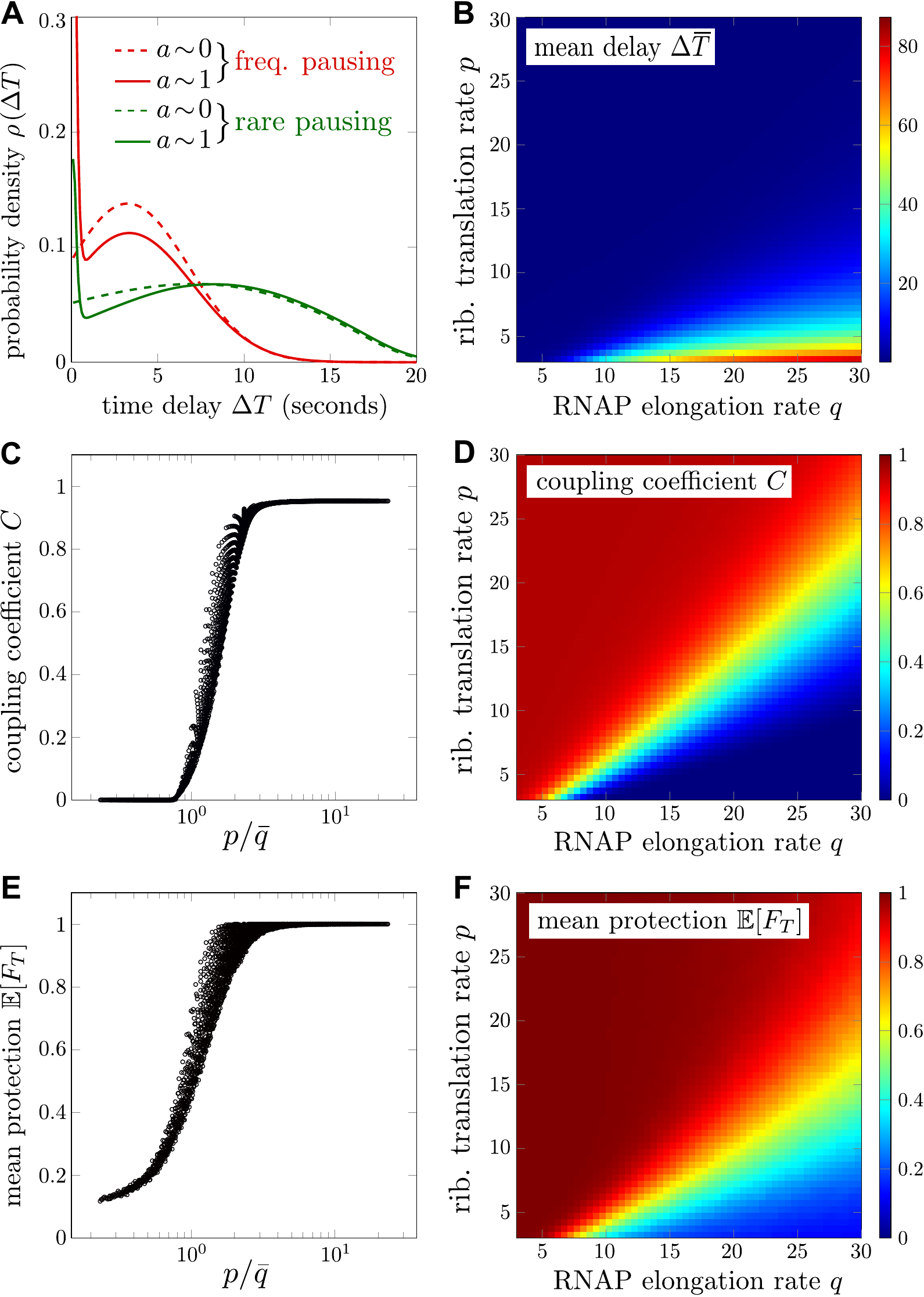}
    \caption{Comparison of different TTC indices.  Common parameters
      for all these plots are $\alpha = 1$/s, $E_+ = 2$, $E_{\rm a}=
      3$, $k_{\rm d}=k_{\rm a}e^{-E_{\rm a}}$, $\ell=4$, $L=335$, and
      $k_{\rm a}=100$/s unless stated otherwise. (A) The delay-time
      distribution $\rho(\Delta T)$ calculated under different
      parameter regimes using Algorithm \ref{alg:update} (see SI).
      Here, $p=12$ codons/s, $q=30$ codons/s, $k_{+}=0.4$/s,
      $k_{-}=0.3$/s for rarely pausing RNAPs (red curves), and
      $k_{+}=4.0$/s, $k_{-}=3.0$/s for frequently pausing RNAPs (blue
      curves).  Fast-binding ($a\sim 1$, $k_{\rm a} = 100$/s, $k_{\rm
        d} = k_{\rm a}e^{-3}$) and no-binding ($a=0$, $k_{\rm a}=0$)
      cases are indicated by solid and dashed curves,
      respectively. (B) Mean delay $\Delta\overline{T}$ as a function
      of $p$ and $q$. (C) The direct coupling coefficient $C$ as a
      function of the relative velocity $p/\bar{q}$. Each point
      represents $C$ evaluated at specific values of $(p,q)$, each
      chosen from all integers between $3$ and $27$ codons/s. (D)
      Heatmap of $C(p,q)$. (E) Values of $\mathbb{E}[F_T]$, each
      derived from 1000 kinetic Monte-Carlo (kMC) trajectories,
      plotted against $p/\bar{q}$. (F) The heatmap of
      $\mathbb{E}[F_{T}(p,q)]$.}
%
    \label{fig:coupling-indices}
\end{figure}

We evaluate our stochastic model to provide quantitative predictions
for the coupling indices, $\rho(\Delta T)$, $C$, $\mathbb{E}[F_{T}]$. The results
are summarized in Fig.~\ref{fig:coupling-indices}.

\paragraph*{Limitations of mean delay time.}
Fig.~\ref{fig:coupling-indices}(A) shows delay-time distributions for
various parameter sets and reveals subtle differences in the kinetic
consequences of coupling.  Without molecular coupling $(k_{\rm a}=0)$,
the distribution has a single peak around the mean delay time. With
molecular coupling, the distribution can exhibit two peaks with one at
$\Delta T=0$. This short-time peak reflects trajectories that
terminate as a bound ribosome-RNAP complex.  These finer structures in
$\rho(\Delta T)$ cannot be resolved by evaluating only the mean delay
time. Fig.~\ref{fig:coupling-indices}B plots the mean delay
$\Delta\overline{T}$ as a function of $p$ and $q$.  For our chosen
parameters, in particular $k_{\rm a} = 100$ s$^{-1}$ and $k_{\rm d} =
k_{\rm a}e^{-3}$, we see that $\Delta \overline{T}$ is rather
featureless, with a significant delay arising only for small
$p$. Thus, the mean delay time provides little information about the
details of TTC.


\paragraph*{Coupling coefficient.}
From an effective velocity argument (see Appendix~\ref{MFT} in the
SI), we approximate the criterion for coupling in terms of

\begin{equation}
  \frac{p}{\bar{q}} \equiv  \frac{p}{q}\frac{k_- + k_+}{k_{+}},
  \label{eqn:determinant}
\end{equation}
where $\bar{q}$ is the average pausing-adjusted RNAP transcription
rate $\bar{q} \equiv qk_{+}/(k_{+}+k_{-})$.
%
%
%
This dimensionless ratio $p/\bar{q}$ is a key indicator of the overall
level of coupling possible. If $p/\bar{q}>1$, the speed of the
ribosome exceeds the average speed of the RNAP, allowing them to
approach each other and potentially form a collided expressome. If
$p/\bar{q}<1$, the ribosome speed is slower than the average RNAP
speed and the system can at most be only transiently coupled. It turns
out that the coupling coefficient $C$ is mostly determined by
$p/\bar{q}$ alone, particularly if all other parameters are kept
fixed. Essentially, the transition to a coupled system (large $C$) is
predicted when $p/\bar{q} \gtrsim 1$. In
Fig.~\ref{fig:coupling-indices}C, we find the values of $C$ for
multiple values of $p$ and $q$ [each dot corresponds to each $(p,q)$
pair], and plot them as a function of $p/\bar{q}$, with
$k_{+}/(k_{+}+k_{-}) \approx 0.43$.  The mean values of $C$ as a function
of $p$ and $q$ are plotted in Fig.~\ref{fig:coupling-indices}D and are
qualitatively distinct from the mean times shown in (B).

\paragraph*{Fraction of time protected.} Each point in 
Fig.~\ref{fig:coupling-indices}E indicates the mean value $F_{T}$,
$\mathbb{E}[F_{T}]$, for different values of $p$ and $q$, arranged
along values of $p/\bar{q}$.  Each mean value $\mathbb{E}[F_{T}(p,q)]$
was computed from averaging protected-time fractions $F_{T}$
(Eq.~\ref{eqn:protected_1}) from 1000 simulated trajectories. As
expected, $\mathbb{E}[F_{T}(p,q)]$ increases linearly with ribosome
translation rate $p$ until saturation to above
$\mathbb{E}[F_{T}]\gtrsim 0.9$ for $p\gtrsim 22$ codons/s.

Comparing $C$ and $\mathbb{E}[F_T]$ from
Figs.~\ref{fig:coupling-indices}D and F, we find that $C$ and
$\mathbb{E}[F_T]$ are qualitatively similar across various values of
$p$ and $q$, although in general we find $\mathbb{E}[F_{T}] \gtrsim
C$.  The transition from low to high values occurs at lower values of
$p$ for $\mathbb{E}[F_{T}]$ since the condition for protection ($d
\leq \ell_{\rm p}$) is not as stringent as that for $C=1$ ($d \leq
\ell$ \textit{and} binding). Thus, there can be value of $(p,q)$ for
which $C(p,q)$ is small but $\mathbb{E}[F_{T}(p,q)]$ is close to one.
%

The similarity between $\mathbb{E}[F_T]$ and $C$ is restricted to the
dependence on $p$ and $q$. The coupling and the protection time
fraction may respond to changes in other parameters in drastically
different ways.  For example, $C$ is nonzero only if molecular binding
is present, rendering it sensitive to $k_{\rm a}, k_{\rm d}$. However,
$F_T$ directly measures the dynamics of TTC and does not depend on
actual molecular coupling, so it will be relatively insensitive to
$k_{\rm a}, k_{\rm d}$, particularly when $p$ is large. Thus, $F_{T}$
may be a better index if we wish to quantify functional consequences
of TTC.  The standard deviations of the simulated $F_{T}$ values are
typically large and approximately
$\sqrt{\mathbb{E}[F_{T}]\big(1-\mathbb{E}[F_{T}]\big)}$ (shown in
Appendix \titleref{variations} of the SI), limiting the suitability of
the mean protected-time fraction as a robust metric.
\begin{figure*}[!tbh]%
    \centering
    \includegraphics[width=5.9in]{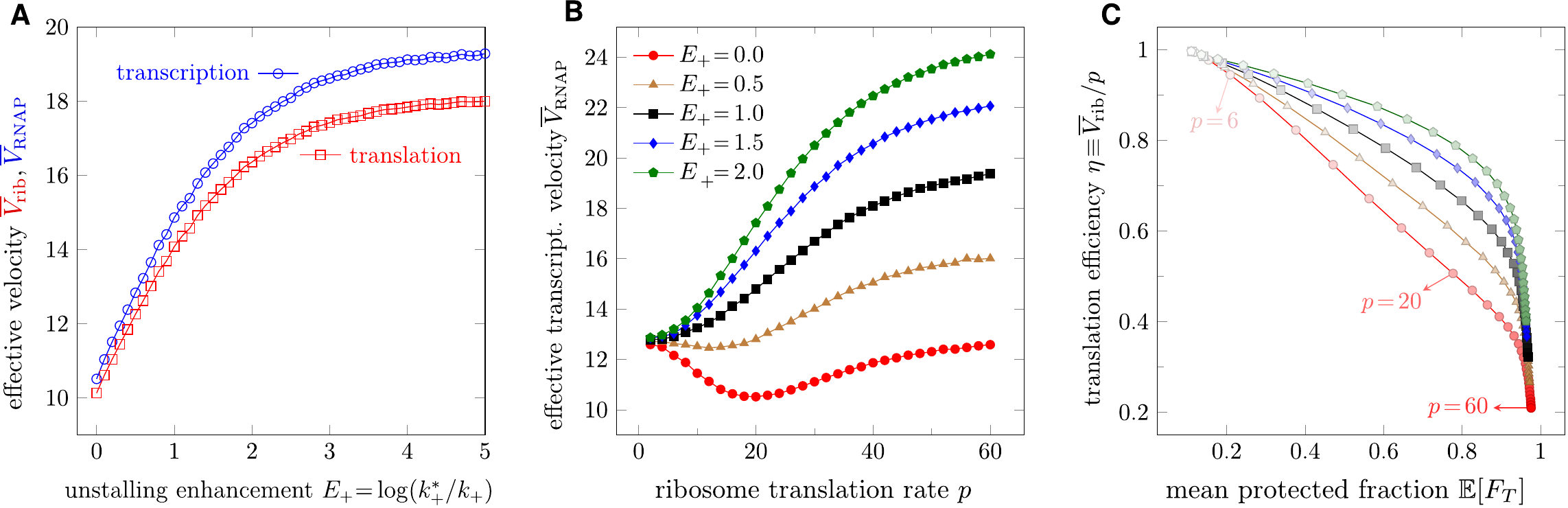}
        \caption{Slowdown induced by molecular coupling. The common
      parameters used are the same as those in
      Fig.~\ref{fig:coupling-indices}.  (A) Effective velocities as a
      function activation energy reduction $E_{+}=\log(k_+^*/k_+)$ in
      ribosome-induced RNAP unstalling, $p= 20$ codons/s and $q=30$
      codons/s. (B) Effective velocity of RNAP as a function of the
      free ribosome translation rate $p$. (C) The trade-off between
      translation efficiency and mean fraction of time protected
      $\mathbb{E}[F_{T}]$.  The efficiency $\eta \equiv
      \overline{V}_{\rm rib}/p$ is defined by the ratio of the mean
      ribosome speed to the translation rate of an isolated ribosome.
      In (B) and (C), $q=30$ codons/s. The variances (not shown) for
      the plotted quantities are large, typically overlapping the
      mean-value curves in (A) and (B).}
     \label{fig:result_2_slow}
\end{figure*}
      



\subsection*{Binding-induced slowdown}

Traditionally, TTC has been invoked as a mechanism for maintaining
RNAP processivity by rescuing RNAP from paused states.  However,
\emph{in vivo} experiments by Kohler et al.~\cite{Kohler2017apr}
reported that when translation is inhibited, the $\Delta \alpha
\textrm{CTD}$ mutant in which RNAP \textit{does not} associate with
ribosome exhibited faster proliferation than that of wild-type RNAP
that can associate with ribosomes. Coupled transcription through
ribosome-RNAP association may give rise to \textit{slower}
transcription.  Thus, TTC may play dual roles of speeding up and
slowing down transcription, depending on conditions.  Through our
model, we will explain the major mechanism of, and limits to,
TTC-induced slowdown of transcription.

\paragraph*{Unstalling rate $k_{+}^{*}$ dictates ribosome efficiency.} 
The principal factor that influences the overall velocity
$\overline{V}$ of a coupled expressome is the interplay between two
antagonistic mechanisms: ribosome-mediated dislodging of an adjacent
stalled RNAP and bound-ribosome deceleration of the RNAP. When the
reduction in activation free energy of unstalling,
$E_{+}=\log(k_{+}^{*}/k_{+})$, is large, the ribosome is less likely
to be impeded by a stalled RNAP.  Fig.~\ref{fig:result_2_slow}A plots
the effective velocities $\overline{V}_{\rm rib}$ and
$\overline{V}_{\rm RNAP}$ as a function of $E_{+}$ and illustrates the
increases in overall speed when the ribosome is more effective at
dislodging a stalled RNAP (higher $E_{+}$).

The decrease in the velocity of a coupled processing RNAP is primarily
determined by the ribosome translation speed $p$. For different values
of $E_+=\log(k_+^*/k_+)$, the dependence of $\overline{V}_{\rm RNAP}$
on $p$ can be quite different, as is shown in Fig
\ref{fig:result_2_slow}B. For large $E_{+}$, when the ribosome
efficiently pushes stalled RNAPs, increasing $p$ allows the ribosome
to more frequently abut the RNAP and dislodge it, leading to faster
overall transcription. However, for inefficient unstalling (small
$E_{+}$), we see that faster ribosomes can \textit{decrease}
$\overline{V}_{\rm RNAP}$. This feature arises because for inefficient
unstalling, a larger $p$ increases the fraction of time the ribosome
and RNAP are bound $(a=1)$, allowing a binding-induced slowdown
to arise more often. Besides $E_{+}$, the emergence of a decreasing
transcription velocity $\overline{V}_{\rm RNAP}$ with increasing
ribosome translation rate $p$ depends intricately on factors such as
$\ell$, $k_{\pm}$, $k_{\rm a}, k_{\rm d}$ and arises only if $k_{\rm
  a}/k_{\rm d}$ is sufficiently large and $\ell$ is not too large.



Although the decrease in $\overline{V}_{\rm RNAP}$ is not large, it
certainly suggests that increasing $p$ under small $E_+ \leq 0.5$ is
not advantageous. This observation motivates us to define a
translation efficiency as the ratio of the effective ribosome speed
$\overline{V}_{\rm rib}$ to its unimpeded translation speed $p$:
${\eta} \equiv \overline{V}_{\rm rib}/p$.  The loss $1-\eta$ measures
how much a ribosome is impeded due to its interactions with the RNAP.
As $p$ is increased, we find trajectories that display a trade-off
between translation efficiency and protected time.  Higher $p$ leads
to more proximal ribosomes and protected RNAP at the expense of
translation efficiency $\eta$.  Fig.~\ref{fig:result_2_slow}C shows
that the decrease in unstalling activation energy $E_+$ affects this
level of trade-off. For large $E_+$, increasing $p$ can speed up
ribosomes beyond the velocity determined by $\bar{q}$ so that $\eta$
decreases more slowly than $\sim 1/p$. At the same time, the system is
only slightly less coupled, leading to a subtle decrease in $\mathbb{E}[F_{T}]$.
In the end, larger $E_{+}$ leads to a higher $\eta$ \textit{versus}
$\mathbb{E}[F_{T}]$ curve.
  
Low $\eta$s are likely selected against since a cell would be
expending more resources than necessary to maintain high levels of
tRNA and other translation factors. An potentially optimal setting may
be to maintain $p\simeq \bar{q}$, which is the minimally sufficient
velocity to keep the RNAP protected. This intermediate choice of $p$
for the ribosome may explain the recent observations that slower
ribosomes did not appreciably slow down transcription \cite{Zhu2019}
or prevent folding of specific mRNA segments \cite{Chen2018}.

\paragraph*{Limits of binding-induced slowdown.}
In cases where $\bar{q} < p <q$ and $L \rightarrow \infty$, the mean
velocity conditioned on coupling ($a=1$) can be estimated in the
strong binding ($k_{\rm a}/k_{\rm d} \gg 1$), steady-state limit (see
Eq.~\ref{eq:velocity_case_b} in Appendix \ref{MFT} of the SI):
\begin{equation}
  \overline{V}_{\rm RNAP} \approx q \left[\frac{(q/p)^{\ell}-1}{(q/p)^{\ell+1}-1}\right] 
                  \frac{k_{+}^{*}}{k_{+}^{*}+k_{-}}.
  \label{eq:coupled_velocity}
\end{equation}
%
%
For large $\ell \geq 1$ and sufficiently large $q/p$, the term
$\left[\frac{(q/p)^{\ell}-1}{(q/p)^{\ell+1}-1}\right]$ is
approximately $p/q$, and lower bounds for
$\overline{V}_{\rm RNAP}(a=1)$ are

\begin{equation}
\overline{V}_{\rm RNAP} \geq \frac{pk_{+}}{k_{+}+k_{-}} \geq
q\left(\frac{k_{+}}{k_{+}+k_{-}}\right)^{2}.
%
\label{eq:coupled_velocity_lower_bound}
\end{equation}
The first equality holds when $k_{+}^{*}=k_{+}$, and the second
equality holds when $p=\bar{q}$. We conclude that the maximum slowdown
induced by binding is essentially limited by the slowdown of RNAP due
to transcriptional road blocks.  The latter plays a fundamental role
in the significantly slower rate of mRNA transcription relative to
rRNA transcription.

\subsection*{Testing the molecular coupling hypothesis}

%
%

\begin{figure*}[tbh!]
\centering
\includegraphics[width=6.1in]{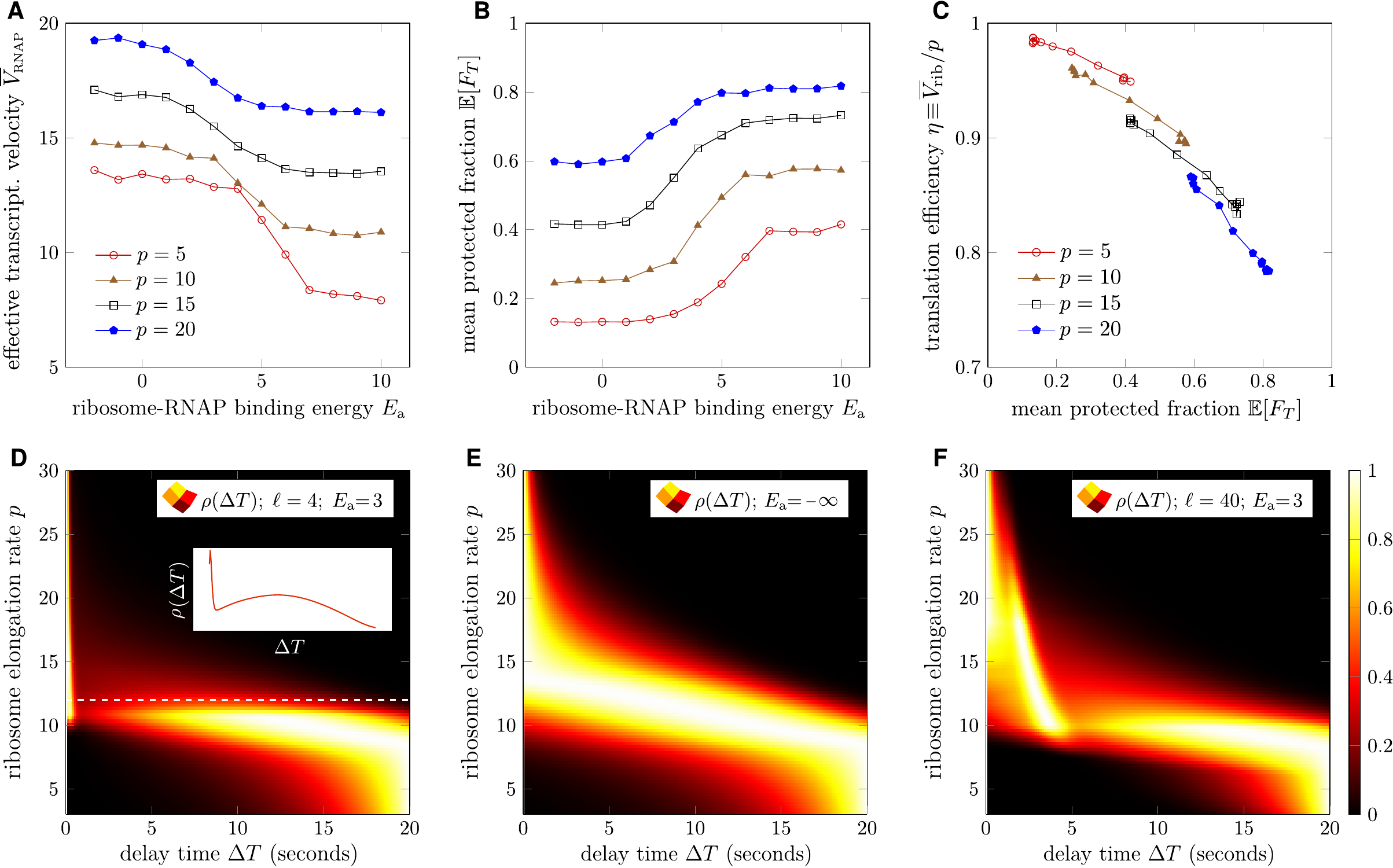}
\caption{Effects of molecular coupling. For those parameters not
  varied, we use the same values used to generate
  Figs.~\ref{fig:coupling-indices} and \ref{fig:result_2_slow}. (A)
  For $\ell = 4$, the effective transcription velocity
  $\overline{V}_{\rm RNAP}$ as a function of binding-energy depth
  between the ribosome and RNAP $E_{\rm a}$.  (B) Mean protected-time
  fraction $\mathbb{E}[F_{T}]$ as a function of binding energy depth
  $E_{\rm a}$ ($\ell=4$). (C) The trade-off between efficiency and
  protection for $\ell=4$. (D) Rescaled heatmap of the delay-time
  distribution $\rho(\Delta T)$ as a function of ribosome
  translocation rate $p$. The brightness indicates the relative
  probability, and the inset shows the probability distribution at
  $p=12$ codons/s indicated by the dashed white line.  Here, the
  binding energy $E_{\rm a} = 3$ and $\ell=4$. For $p\approx 9$
  codons/s, $\rho(\Delta T)$ is bimodal in $\Delta T$.  (E) Delay-time
  distribution in the absence of ribosome-RNAP binding ($k_{\rm a} =
  0$). Here, the $\ell$-dependence disappears and $\rho(\Delta T)$ is
  monomodal.  (F) Delay-time distribution for $\ell=40$ and $E_{\rm a}
  = 3$.  Bimodality arises in more than one regime of $p$.}
\label{fig:result_3}
\end{figure*}

Since there has not been direct observation of molecular coupling
\textit{in vivo}, it is informative to compare scenarios that predict
molecular coupling to those that do not.  We now vary the binding
energy $E_{\rm a}$ for different velocity ratios $p/\bar{q}$.  For
$\ell=4$, Fig.~\ref{fig:result_3}A shows $\overline{V}_{\rm RNAP}$ as
a function of $E_{\rm a}$ for various values of $p$. Although higher
$p$ leads to increased $\overline{V}_{\rm RNAP}$, for each value of
$p$, increasing the binding energy increases coupling and leads to
RNAP slowdown.  Both $p$ and $E_{\rm a}$ increase $\mathbb{E}[F_{T}]$
as shown in Fig.~\ref{fig:result_3}B. As different values of $E_{\rm
  a}$ are used, we also find a trade-off between ribosome efficiency
and protection, as shown in Fig.~\ref{fig:result_3}C. In Appendix
\ref{interaction_ell}, we provide additional simulation results that
confirm the $\ell$-dependence in Eq.~\ref{eq:coupled_velocity} and in
$\mathbb{E}[F_{T}]$.

%
%

Our predicted differences in effective velocities $\overline{V}$ and
$\mathbb{E}[F_{T}]$ are probably not significant enough to be easily
distinguished experimentally; thus, we investigate the distribution of
delay times $\rho(\Delta T)$ as $p$ is varied.
Fig.~\ref{fig:result_3}D shows $\rho(\Delta T)$ rescaled so that the
largest value is set to unity for easier visualization. We see that
for intermediate values of $10\lesssim p\lesssim 13$, $\rho(\Delta T)$
can be bimodal.  Fig.~\ref{fig:result_3}E depicts a single-peaked
$\rho(\Delta T)$ when coupling is completely turned off by setting
$k_{\rm a} = 0, (E_{\rm a} = -\infty)$. In this case, $\ell$ is
irrelevant. Fig.~\ref{fig:result_3}F shows the rescaled $\rho(\Delta
T)$ in the presence of coupling ($E_{\rm a} = 3$) for $\ell=40$. Here,
there are two regimes, $8\lesssim p\approx 12$ and $18\lesssim
p\lesssim 24$, that exhibit bimodality.
%
%

\subsection*{Genome-wide variability of coupling}

We have so far assumed all parameters are homogeneous along the
transcript and time-independent. However, a cell is able to
dynamically regulate the transcription and translation of different
genes by exploiting the transcript sequence or other factors that
mediate the process. Such regulation can be effectively described
within our model by varying its parameters in the appropriate way.


\paragraph*{Regulation of RNAP pausing.}
The RNAP pausing rate $k_{-}$ is one parameter that can be modulated
by specific DNA sequences and other roadblocks along the gene
\cite{Komissarova1997,Epshtein2003may,Nudler2009,John2000}.  There is
evidence that consensus pause sequences are enriched at the beginning
of genes \cite{Hatoum2008,Larson2014may}.  In addition to leading
ribosomes, a trailing RNAP can also push the leading RNAP out of a
paused state by increasing $k_{+}$, much like ribosomes
\cite{Epshtein2003may,Zuo2022}. Even if $k_{+}$ and $k_{-}$ are varied
in our model, the overall predicted performance regimes of the system
are still delineated by values of $p/\bar{q}$, and the effective
transcription velocity can still be predicted by
Eq.~\ref{eq:coupled_velocity}.

\paragraph*{Effects of translation initiation rates.}
Translation initiation is another process that can be altered by the cell
through, \textit{e.g.}, initiation factors that modulate the
initiation rate $\alpha$ \cite{Chou2003}.  Genome-wide analysis
reveals that translation initiation times in E. coli are highly
variable, ranging from less than 1 second to more than 500 seconds
\cite{Siwiak2013sep,Shaham2017nov}.

\begin{figure*}[htb!]
\begin{center}
\includegraphics[width=5.8in]{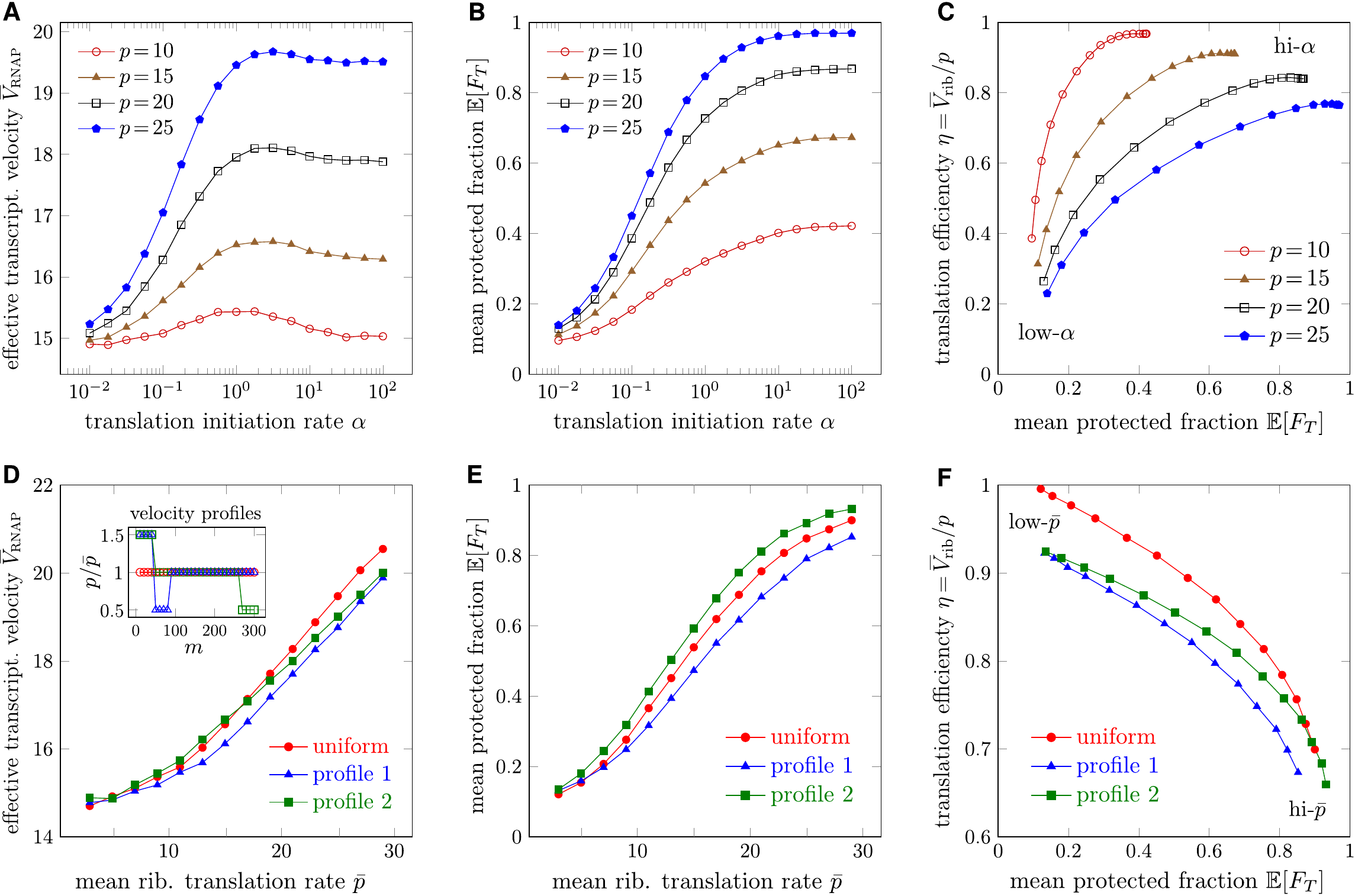}
  \caption{TTC and performance under genomic variability.  Again, we
    use the standard set of fixed parameter values as in
    Figs.~\ref{fig:result_2_slow} and \ref{fig:result_3}.  (A)
    Effective transcription velocity $\overline{V}_{\rm RNAP}$ as a
    function of the translation-initiation rate $\alpha$.
    $\overline{V}_{\rm RNAP}$ initially increases as $\alpha$ is
    increased until about $\alpha \simeq 1$ s $^{-1}$, after which
    $\overline{V}_{\rm RNAP}$ decreases slightly as $\alpha$ is
    further increased.  (B) Mean protected fraction
    $\mathbb{E}[F_{T}]$ as a function of the translation-initiation
    rate $\alpha$. (C) Efficiency \textit{versus} protection fraction
    as $\alpha$ is varied. Larger $\alpha$ contributes to both
    efficiency and protection.  As $\alpha$ is increased, the system
    spends more time protected.  Since $\overline{T}_{\rm rib}$
    includes the ribosome initiation time, it decreases as $\alpha$ is
    increased, leading to an increased $\overline{V}_{\rm rib}$ and
    $\eta$.  (D) and (E) $\overline{V}_{\rm RNAP}$ and
    $\mathbb{E}[F_{T}]$ as a function of the mean $\bar{p}$ under
    three different translocation rate profiles $p_{m}$. (F) The
    $\eta$-$\mathbb{E}[F_{T}]$ trade-off plot for three different
    profiles as the mean value $\bar{p}$ is varied.  The inset in (D)
    illustrates the three translation-rate profiles. The overall
    performance of profile 1 suffers because of the slowdown following
    the initial fast translation.  For $\bar{p} \leq 15$ codons/s,
    profile 2 has a higher $\overline{V}_{\rm RNAP}$ compared with the
    uniform profile. When $p \geq 15$ codons/s, the uniform profile
    still has a higher $\overline{V}_{\rm RNAP}$. For all values of
    $p$, profile 2 has a higher $F_T$ than the uniform profile because
    of its higher initial speed.}
  \label{fig:result_4}
\end{center}
\end{figure*}

As shown in Figs.~\ref{fig:result_4}A-C, varying the translation
initiation rate $\alpha$ straightforwardly affects TTC.  As indicated
in (A), the predicted $\overline{V}_{\rm RNAP}$ at $\alpha \simeq 1
\textrm{s}^{-1}$ is preserved across different values of $p$.  Slower
translation initiation results in larger initial separations $n_{0}$,
decreasing the overall fraction of protected times, as shown in (B)
and (C).  To mitigate large initial distances $n_{0}$ and lower
likelihood coupling due to slow initiation, RNAP pausing occurs more
often at the start of the gene to allow time for a slow-initiating
ribosome to catch up. Thus, delayed ribosome initiation and early RNAP
pausing are two ``opposing'' processes that can regulate coupling and
efficiency, especially for short genes.


\paragraph*{Ribosome translocation-rate profiles.}
Although we have thus far assumed uniform ribosome translocation
rates, it is known that codon bias and tRNA/amino acid availability
can locally affect ribosome translocation
\cite{Lakatos2004,Klumpp2012}.  Snapshots of ribosome positions along
transcripts have been inferred from ribosome profiling
experiments. After imposing a stochastic exclusion model
\cite{MacDonald1969}, Khanh et al. \cite{DaoDuc2018jan}
reconstructed position-dependent ribosome translocation rates
$p_{m}$. They found that hopping rates $p_{m}$ are larger near the 5'
end and decreases towards the 3' end. Although they reconstructed the
entire genome-wide $p_{m}$ profile, translocation rates are
gene-dependent, so we will propose and test simple profiles $p_{m}$.

To qualitatively match the inferred profile \cite{DaoDuc2018jan}, we
define profile 1 by increasing $p$ by 50\% for the first 40 codons,
and decreasing it by 50\% for the second 40 codons. The rest of the
transcript retains the constant baseline value of $p$. Profile 2 is
similarly defined except that instead of being the second group of 40
codons, the speed across the last 40 codons is decreased.  We compared
the performance of the three different profiles in
Fig.~\ref{fig:result_4}D-F as a function of the mean translation rate
$\bar{p} \equiv L^{-1}\sum_{m=1}^{L}p_{m}$.  In the low-speed regime
$\bar{p} \lesssim 10$ codons/s, higher starting $p_{m}$ values promote
ribosome-RNAP interactions, leading to a slightly higher effective
transcription velocity $\overline{V}_{\rm RNAP}$ and higher mean
protected fraction $\mathbb{E}[F_{T}]$. However, in profile 1, the
subsequent decrease in $p_{m}$ under strong coupling is sufficient to
induce slowdown of RNAP. This nonmonotonic effect is weaker in profile
2 because by the later time that translational slowdown occurs, the
machines are further apart (and less likely to be bound) since they
are further removed from the common initial high-$p$ region.
Fig.~\ref{fig:result_4}F shows that profile 2 provides the best
protection, but increasing the likelihood of coupling means that
profiles 1 and 2 are more likely to be impeded by stalled RNAP,
leading to slightly lower ribosome efficiency $\eta$.


\section*{Summary and Conclusions}

We have presented a detailed stochastic model of
translation-transcription coupling (TTC). The continuous-time
discrete-state model tracks the distance between the leading ribosome
and the RNAP and assumes they sterically exclude each other along the
nascent mRNA transcript. All current experimental understanding of
interactions between RNAP and ribosome, including ribosome initiation,
RNAP pausing, and direct ribosome-RNAP association have also been
incorporated. Our model exhibits a number of rich features that depend
on the interplay of these intermediate mechanisms.

To quantitatively investigate the predictions of our model, we
constructed three different metrics to quantify TTC, the delay-time
probability distribution $\rho(\Delta T)$, the probability $C$ that
the ribosome and the RNAP are in a bound state $(a=1)$ at termination,
and the fraction of time $F_{T}$ that the ribosome and the RNAP are
proximal over the entire transcription-translation process.  $F_{T}$
is a measure of protection against binding of termination
proteins. These metrics were computed or simulated under different
model parameters.
%
%
Specifically, since a bound RNAP at distance $\ell$ from the trailing
ribosome needs to first detach before $d=\ell$ can be increased, the
$d \leq \ell$ states shown in Figs.~\ref{fig:mechanism} and
\ref{fig:distance_state_space} form an effective attractive well that
tethers RNAP to ribosome.  By allowing direct ribosome-RNAP binding,
we find that this effective attraction zone can allow a slower
ribosome to dynamically hold back bound RNAP, leading to decreased
$\overline{V}_{\rm RNAP}$.

Qualitatively, our model predicts two different regimes of TTC that
appear to be consistent with observations.  One limit can arise when
$E_{\rm a}$ is large, resulting in close proximity and strong
molecular coupling that may lead to slowdown of RNAP, while the other
arises when $E_{\rm a}$ is small leading to intermittent contacts and
perhaps modest speed up of pausing RNAPs. Besides $E_{\rm a}$, our
model suggests that $\ell$, $\alpha$ and $p/\bar{q}$ also control
which type of TTC arises. Across different genes, $E_{\rm a}$ and
$\ell$ are expected to be unchanged, but variations in $p/\bar{q}$
(and to some degree $\alpha$) can affect the balance between these
qualitative models of TTC. For example, it may be advantageous to
produce housekeeping genes as rapidly as possible through strong
coupling, while for other genes, translation efficiency may be more
important and achieved by weak coupling, at the expense of protection
and speed. Our model reconciles these two limits under a unified model
that distinguishes the gene-specific parameters that can modulate the
form of TTC.

If TTC is mediated by, say, NusG, the effective binding energies
associated with the ribosome-NusG-RNAP complex will be critical. While
these binding energies are unknown, NusG-mediated TTC can form a
larger expressome complex, allowing for more confined mRNA which we
can take to be $\ell\sim 40$.  Compared to direct TTC with $\ell = 4$,
the larger value of $\ell$ in NusG-mediated interactions can also
yield higher efficiency $\eta$ and protection $F_{T}$. By controlling
NusG availability, the coupling can dynamically switch between large
and small complexes. Although NusG-mediated TTC is qualitatively
similar to direct short-ranged TTC in that both scenarios can exhibit
bimodal delay-time distributions $\rho(\Delta T)$, a dynamically
varying $\ell$ can further fine-tune ribosome induced slowdown.

%

%

Although experimental verification of direct molecular binding during
transcription is lacking, our model reveals that a bimodal time-delay
distribution when $p\approx\bar{q}$ is a hallmark of molecular
association. Protocols such as single-molecule DNA curtains may
provide information on the effective and instantaneous velocity of
RNAP under different translation elongation rates.  By comparing
velocities to theoretical predictions, it may be possible to infer the
unstalling enhancement $E_+$. Finally, FRET experiments or
super-resolution imaging may shed light on macromolecular-level
ribosome and RNAP dynamics \cite{Stasevich2016}.  Our model can guide
how \textit{in vitro} measurements can be designed and used to
reconstruct delay-time distributions $\rho(\Delta T)$, coupling
coefficients $C$, protected-time fractions $F_T$, and efficiencies
$\eta$.


\section*{Author Contributions}
XL and TC devised and analyzed the model and wrote the paper.  XL
developed the computational algorithms and performed the numerical
calculations and kinetic Monte Carlo simulations.

\section*{Declaration of Interest}
The authors declare no competing interests.

\section*{Acknowledgments}
This work was supported by grants from the NIH through grant
R01HL146552 and the NSF through grant DMS-1814364 (TC).


\bibliography{library}

\clearpage

\renewcommand{\theequation}{S\arabic{equation}}
\renewcommand{\thefigure}{S\arabic{figure}} 
\renewcommand{\thesubsection}{S\arabic{subsection}}
\setcounter{subsection}{0}   
\setcounter{figure}{0}   
\setcounter{equation}{0}  
\setcounter{page}{1}



\onecolumn

\section*{Supplementary Information: Mathematical Appendices}

\subsection{Stochastic Simulations} 
\label{SS}

Although numerical and analytic evaluation of the master equation
associated with our stochastic model is possible in some limits,
certain quantities such as the fraction of protected time $F_{T}$ are
most easily evaluated via Monte-Carlo simulation.  We employed an
event-based kinetic Monte Carlo algorithm to simulate trajectories of
our full model.  The Gillespie \cite{Gillespie1977dec} or
Bortz-Kalos-Lebowitz algorithm \cite{Bortz1975} first finds all the possible
reactions and their rates. Then, one randomly chooses, with
probability weighted by all the reaction rates, a reaction to fire.
An independent random number is again drawn from the exponential
distribution with rate equal to the total reaction rates. The relevant
code is available at \url{https://github.com/hsianktin/ttc}.

\subsection{Master equation}
\label{numerical}

The probability of a state $\omega=(m,n,a,b)\in \Omega$ at
time $t$ is defined by  $\mathbb{P}_t(m,n,a,b)\equiv \mathbb{P}[\omega_t=(m,n,a,b)]$.
Here, $\Omega$ is the sample space of all allowable $(m,n,a,b)$.
%
%
%
Because the $(a,b)\in \left\{ 0,1 \right\}^2$ contains only four
components, we can flatten the four-component probability
$\mathbb{P}_{t}$ by introducing the $(m,n)$-dependent probability
vectors $\textbf{P}(m,n) = (P_{0}(m,n), P_{1}(m,n), P_{2}(m,n),
P_{3}(m,n))^{T}$ in which the components describe the probabilities
associated with the internal $(a,b)$ configurations when the ribosome
and the RNAP are at positions $(m,n)\in \Omega_{mn}$:

\begin{itemize}
	\item $P_{0}$: ribosome and RNAP are unassociated and both processing ($a=0,b=0$)
	\item $P_{1}$: processing ribosome, but paused, unassociated RNAP ($a=0,b=1$)
	\item $P_{2}$: associated ribosome/RNAP, both in processing states ($a=1,b=0$)
	\item $P_{3}$: associated ribosome/RNAP, paused RNAP ($a=1,b=1$)
\end{itemize}
The last two states can only arise when the ribosome and the RNAP are
within the interaction range $d \equiv n-m \leq \ell$.

The transition matrix describing transitions among elements of the $4
\times 1$ probability vector $\mathbf{P}$ are organized in terms of $4\times 4$
matrices ${\textbf{p}}_{m,n}$, ${\textbf{q}}_{m,n}$, and $\textbf{k}_{m,n}$

%
\begin{equation}
\begin{aligned}
{\textbf{p}}_{m,n} & = \left(\begin{array}{cccc} p_{m} & 0 & 0 & 0 \\
0 & p_{m} & 0 & 0 \\
0 & 0 & p_{m} & 0 \\
0 & 0 & 0 & p_{m} \end{array}\right),\,\,\,
{\textbf{q}}_{m,n}=\left(\begin{array}{cccc} 
q_{n} & 0 & 0 & 0 \\
0 & 0 & 0 & 0 \\
0 & 0 & q_{n}\mathds{1}_{d< \ell} & 0 \\
0 & 0 & 0 & 0\end{array}\right),\,\,\, \textbf{k}_{0,n} = \left(\begin{array}{cccc}
- k_{-} & k_{+} & 0 & 0 \\
k_{-} & - k_{+} & 0 & 0 \\
0 & 0 & 0 & 0 \\
0 & 0 & 0 & 0 \end{array}\right)\\[10pt]
\textbf{k}_{m\geq 1,n}  & = \left(\begin{array}{cccc}
- k_{-} - k_{\rm a}(m,n) & k_{+} & k_{\rm d} & 0 \\
k_{-} &  - k_{+}-k_{\rm a}(m,n) & 0 & k_{\rm d} \\
k_{\rm a}(m,n) & 0 & - k_{-} - k_{\rm d} & k_{+} \\
0 & k_{\rm a}(m,n) & k_{-} & -k_{+} - k_{\rm d}\end{array}\right),\\[10pt]
\textbf{k}_{n,n} & = \left(\begin{array}{cccc}
- k_{-} - k_{\rm a} & k_{+}^{*} & k_{\rm d} & 0 \\
k_{-} &  - k_{+}^{*}-k_{\rm a} & 0 & k_{\rm d} \\
k_{\rm a} & 0 & - k_{-} - k_{\rm d} & k_{+}^{*} \\
0 & k_{\rm a} & k_{-} & -k_{+}^{*} - k_{\rm d}\end{array}\right),
\end{aligned}
\end{equation}
%
%
where ${\textbf{p}}_{m,n}$ and ${\textbf{q}}_{m,n}$ contain processive
ribosome and processive RNAP hopping rates and $\textbf{k}_{m,n}$ is
the transition rate matrix connecting the internal ribosome/RNAP
states.  Here, $\mathds{1}_{z}=1$ if an only if $z$ is satisfied.
Ribosome-RNAP exclusion is imposed via $\textbf{p}_{n,n}=0$ and
ribosome initiation is defined by $p_{m=0}\equiv \alpha$.  The
internal-state conversion rate matrix depends on $(m,n)$ via $k_{\rm
  a}(m,n) = k_{\rm a}\mathds{1}_{d\leq \ell}$.  For simplicity, we
assume the values of the intrinsic kinetic rates $k_{\pm}, k_{\rm
  a,d}$ to be otherwise $(m,n)$-independent (although $p_{m}$ and
$q_{n}$ can still depend on position).  The master equation is then
given by
\begin{equation}
  {\frac{\partial {{\textbf{P}}}(m,n)} {\partial t}} 
={\textbf{p}}_{m-1,n}{{\textbf{P}}}(m-1,n) + {\textbf{q}}_{m,n-1}{\textbf{P}}(m,n-1) 
- ({\textbf{p}}_{m,n}+{\textbf{q}}_{m,n}){\textbf{P}}(m,n)+\textbf{k}_{m,n} \textbf{P}(m,n), 
\,\,\,0\leq m\leq n,
\end{equation}
with boundary conditions
$\textbf{P}(-1,n)=\textbf{P}(m,-1)=\textbf{P}(m,n)|_{m>n}=0$.  For
time-homogeneous problems, we define the time Laplace-transformed
probability vector $\mathcal{L}\{P(m,n,t)\} \equiv
\tilde{\textbf{P}}_{m,n}(s)$, which satisfies
\begin{equation}
 s {\tilde{{\textbf{P}}}}_{m,n} - {\textbf{P}}(m,n, t=0) = 
 {\textbf{p}}_{m-1,n}\tilde{{\textbf{P}}}_{m-1,n} +{\textbf{q}}_{m,n}
\tilde{{\textbf{P}}}_{m,n-1}- ({\textbf{p}}_{m,n}
+{\textbf{q}}_{m,n})\tilde{{\textbf{P}}}_{m,n}
+{\textbf{k}}_{m,n}{\tilde{{\textbf{P}}}}_{m,n}.
\label{master_s}
\end{equation}
We set the initial condition $P_{i}(m,n,t=0) = \mathds{1}_{m,0}
\mathds{1}_{n,1}\mathds{1}_{i,1}$ to describe a ribosome-free system
immediately after a processing RNAP has started transcription.  The
probability of this state is then self determined by
$\tilde{{\textbf{P}}}_{0,1}(s) = {\boldsymbol\gamma}^{-1}_{0,1}(s)
\textbf{e}_1$ where
%
%
$\textbf{e}_1 = (1,0,0,0)^T$ and
${\boldsymbol\gamma}_{0,1}(s)=(s{\bf I}+{\bf p}_{0}+{\bf
  q}_{0,1}-{\bf k}_{0,1})$, where ${\bf I}$ is the 
identity matrix. Starting from this value, we can evaluate
the vector recursion relation in Eq.~\ref{master_s}.  Be defining
${\boldsymbol\gamma}_{m,n}\equiv (s{\bf I}+{\textbf{p}}_{m,n} +
{\textbf{q}}_{m,n} - {\textbf{k}}_{m,n})$ by
${\boldsymbol\gamma}_{m,n}$, the recursion relation is simplified to
\begin{equation}
  \tilde{{\textbf{P}}}_{m,n} = {\boldsymbol\gamma}_{m,n}^{-1} 
\left[{\textbf{p}}_{m-1,n} \tilde{{\textbf{P}}}_{m-1,n} 
+ {\textbf{q}}_{m,n-1}\tilde{{\textbf{P}}}_{m,n-1} \right].
  \label{eq:recursion}
\end{equation}
This structure allows us to combine terms into overall 
transition kernels
\begin{equation}
\textbf{j}^{m,n}_{m-1,n}\! ={\boldsymbol\gamma}^{-1}_{m,n} {\textbf{p}}_{m-1,n}, \quad
\textbf{j}^{m,n}_{m,n-1}\! ={\boldsymbol\gamma}^{-1}_{m,n}{\textbf{q}}_{m,n-1}.
\label{eq:transition}
\end{equation}

If $\Theta$ is the set of all possible paths
$\theta=(\theta_1,\ldots,\theta_{m+n-1})$ connecting
$(0,1)$ and $(m,n)$. The probability $\tilde{P}_{m,n}$ can be recursively found by
a weighted sum of all possible paths from $(0,1)$ to $(m,n)$:
\begin{equation}
\tilde{{\textbf{P}}}_{m,n} = \sum_{\theta\in \Theta}
 \prod_{i=2}^{m+n-1} \textbf{j}^{\theta_{i}}_{\theta_{i-1}}
 \tilde{{\textbf{P}}}_{0,1} 
\label{eq:path}
\end{equation}
Since ${\textbf{p}}, {\textbf{q}}$, and $\textbf{k}$ are pairwise
commutative, using Eq.~\ref{eq:transition} in Eq.~\ref{eq:path}, we
find
\begin{equation}
  \tilde{{\textbf{P}}}_{m,n} =  \sum_{\theta\in \Theta}
  {{\textbf{p}}^m {\textbf{q}}^{n-1}}{\prod_{i=2}^{m+n-1} 
{\boldsymbol\gamma}_{\theta_i}^{-1}}\tilde{{\textbf{P}}}_{0,1}.
  \label{eq:path_integral}
\end{equation}
The recursion relation Eq. \ref{eq:recursion} can be evaluated
numerically to find $\tilde{{\textbf{P}}}_{m,n}$, while the path
integral Eq.~\ref{eq:path_integral} can be used to approximate
analytic solutions in specific limits. For example, if there is no
ribosome-RNAP binding, and all other parameters are homogeneous,
$\tilde{{{P}}}_{3}=\tilde{{{P}}}_{4}=0$. We can project all parameters
and variables into a two-dimensional subspace supported by $\left\{
\tilde{P}_0, \tilde{P}_1 \right\}$. The only interactions considered
are the volume exclusion effects. In this case, ${\boldsymbol\gamma}$
assumes two possible values. In the interior ($m<n$),
${\boldsymbol\gamma}_{\textrm{in}}= (s{\bf I} + {\textbf{p}} +
{\textbf{q}} - \textbf{k})$ while on the boundary
$\partial\Omega_{mn}\equiv \{(m,n)\in \Omega_{mn}: m=n\}$,
${\boldsymbol\gamma}_{\textrm{ex}}=(s{\bf I} + {\textbf{q}} -
\textbf{k})$.

We can classify different paths $\theta$ by the number of visits $w$
to the boundary before reaching $(m,n)$: $\Theta_w\equiv
\{\|\{\theta_i\}_{i=1}^{\|\theta\|-1} \cap \partial \Omega\|=w\}$.

Eq. \ref{eq:path_integral} is then rearranged to be
\begin{equation}
	\tilde{{\textbf{P}}}_{m,n} = \sum_{w=0}^{\infty} 
  \sum_{\theta\in \Theta_w}
   {\textbf{p}}^m {\textbf{q}}^{n-1}{\boldsymbol\gamma}_{\textrm{ex}}^{-w} 
  {\boldsymbol\gamma}_{\textrm{in}}^{-(m+n-1-w)} \tilde{{\textbf{P}}}_{0,1}
   \label{eq:analytic_solution}
\end{equation}

\paragraph*{Analytic solution for the first passage problem.}
A simpler closed-form analytic solution can be obtained when
considering a first passage problem to the boundary
$\partial\Omega_{mn}$.  If $\omega_t$ denotes the stochastic process
of the TTC problem, we wish to find the probability that the position
$\omega_t$ at time $t$ is $(m,n)$ and that at $T_{\rm b} \geq t$:
$\mathbb{P}(\omega_t=(m,n), T_{\rm b} \geq t)$.

To solve this problem, we use the method of coupling. Consider a
second, absorbing process $\omega'_t$ which is coincidental with
$\omega_t$ up until $T_{\rm b}$, upon which it ceases to evolve. In
other words, $\omega'_t = \omega_{\min\{t,T_{\rm b}\}}$.  For the
$\omega'_{t}$ process, the Laplace-transformed probability satisfies
\begin{equation}
\tilde{{\textbf{P}}}_{m,n} =  \sum_{\theta\in \Theta_0} {\textbf{p}}^m 
{\textbf{q}}^{n-1} {\boldsymbol\gamma}_{\textrm{in}}^{-(m+n-1)} 
\tilde{{\textbf{P}}}_{0,1}.
\label{eq:simplified_probability}
\end{equation}
Note that each term in the summation does not depend on the actual
path $\theta \in \Theta_0$. Therefore, we just need to calculate the
size of $\Theta_0$, which is a generalized problem of finding Catalan
numbers. Obviously, when $m < n-1$, the size is simply the binomial
coefficient $\binom{m+n-1}{m}$. 

To proceed further, we need to further assume $\ell=1$ before
calculating powers of the truncated $2\times 2$ matrices
${\boldsymbol\gamma}_{\textrm{in}}$ by first diagonalizing
\begin{equation}
{\boldsymbol\gamma}_{\textrm{in}} = \begin{pmatrix}
s + p + q + k_{-} & - k_{+} \\ -k_{-} & s+p +k_+
\end{pmatrix} = {\bf V}^{-1} {\bf D V},
\end{equation}
where
\begin{equation}
  \begin{aligned}
\textbf{V} & \displaystyle = \left[\begin{matrix}
  \displaystyle \frac{- k_{-} + k_{+} - q + \delta}{2 k_{-}} 
  &  \displaystyle \frac{- k_{-} + k_{+} - q - \delta}{2 k_{-}}\\[10pt]
  1 & 1\end{matrix}\right], \\[8pt]
\textbf{D} & = \textrm{diag}\left[s+\frac{k_{-}}{2} 
+ \frac{k_{+}}{2} + p + \frac{q}{2} - \frac{\delta}{2},\,
s+\frac{k_{-}}{2} + \frac{k_{+}}{2} + p + \frac{q}{2} + \frac{\delta}{2}\right],\\[8pt]
\delta & \equiv  \sqrt{k_{-}^{2} + 2 k_{-} k_{+} + 2 k_{-} q + k_{+}^{2} - 2 k_{+} q + q^{2}}.
  \end{aligned}
\end{equation}
Then, ${\boldsymbol\gamma}_{\textrm{in}}^n =
\textbf{V}^{-1}\textbf{D}^n \textbf{V}$ for all $n\in \mathbb{N}$.
The ${\textbf{p}}$ and ${\textbf{q}}$ matrices are both diagonal, and
their powers are straightforward to calculate.  Thus, as long as the
combinatoric factors can be calculated, Eq.~\ref{eq:analytic_solution}
and \ref{eq:simplified_probability} can be expressed in analytic
forms.

\subsection{Numerical procedure for conditional distributions}
\label{conditional_distribution} 

We also developed an iterative numerical algorithm for numerically
approximating the probability distribution of the ribosome location
$m$, the RNAP position, the RNAP state $b$, and the ribosome-RNAP
association state $a$.  The algorithm is detailed below.
  \begin{algorithm}
    \caption{Algorithm for updating the conditional distribution}\label{alg:update}
    \begin{algorithmic}[1]
    \Procedure{Update}{}[$\mathbb{P}(m,a,b\vert n)$, tolerance]
    \State $\varepsilon \gets$ tolerance 
    \State $\mathbb{P}(m,a,b\vert n+1) \gets 0, \forall m,a,b$
    
    \While {$\sum_{m,a,b} \mathbb{P}(m,a,b\vert n+1) < 1 - \varepsilon$}
    \For {($m,a,b$, $m',n',a',b'$)}
    \State $\mathbb{P}(m',a',b'\vert n') \gets
    \mathbb{P}(m',a',b'\vert n')+J^{m',n',a',b'}_{m,n,a,b}
    \mathbb{P}(m,a,b\vert n)$
%
%
    \State $\mathbb{P}(m,a,b\vert n) \gets \mathbb{P}(m,a,b\vert
    n)-J^{m',n',a',b'}_{m,n,a,b}\mathbb{P}(m,a,b\vert n)$
%
%
    \EndFor
    \EndWhile
    \State \textbf{return} $\mathbb{P}(m,a,b\vert n+1)$.
    \EndProcedure
    \State
    \Procedure{$J^{m',n',a',b'}_{m,n,a,b}=~$}{}$J(m',n',a',b'\vert m,n,a,b)$
    \State $R_{\textrm{tot}} \gets \sum_{m'',n'',a'',b''} r^{m'',n'',a'',b''}_{m,n,a,b}$
    \State \textbf{return} $\left(r^{m',n',a',b'}_{m,n,a,b}/R_{\textrm{tot}}\right)$
    \EndProcedure
    \end{algorithmic}
  \end{algorithm}
Again, use $\omega_t$ to denote the full state $(m_t,n_t,a_t,b_t)$ of
the system at time $t$.  Let $\tau_0=0$, and recursively define
$\tau_{n}$ as follows:
\begin{equation}
  \tau_{n} = \inf \left\{ t>\tau_{n-1}: \omega_t \neq \omega_{\tau_{n-1}} \right\}
\end{equation}
Let $\textbf{w}_{n} = \omega_{\tau_n}$. Then, $\textbf{w}_{n}$ is a discrete Markov
chain on the same state space $\Omega$ as $\omega_t$ and satisfies
\begin{equation}
  \mathbb{P}(\textbf{w}_{n}|\textbf{w}_{n-1}) 
= \frac{r(\textbf{w}_{n}|\textbf{w}_{n-1})}{\sum_{\omega\in \Omega} 
r(\omega|\textbf{w}_{n-1})}, \quad \forall\, n \geq 1
\end{equation}
where $r(\omega|\textbf{w}_{n-1})$ is defined in
Eqs.~\ref{eqn:initiation} -\ref{eqn:restarting}.

In order to find the distribution of ribosome positions $m$ upon
completion of transcription at time $T_{\rm RNAP}$, we first define
the stopping times $t_k$ as the discrete-time analog of $T_{k}$ such
that ${\bf w}_{t_{k}} = \omega_{T_{k}}$.
%
%
%
Then, we use the shorthand notation $\mathbb{P}(m,a,b\,\vert\,
t=T_{n})$ defined by
\begin{equation}
  \mathbb{P}(m,a,b\,\vert\,t=T_{n}) = \mathbb{P}(\omega_{T_{n}}=(m,n,a,b))
\mathbb{P}(\textbf{w}_{t_n}=(m,n,a,b)).
  \label{eqn:definition_conditional_distribution}
\end{equation}
%
%
Upon defining $t^{(k)}_{n} \coloneqq \min \{(t_{n} + k), t_{n+1}\}, \,
\lim_{k\rightarrow \infty} t^{(k)}_{n} = t_{n+1}$
%
%
guarantees the pointwise convergence of 
$\textbf{w}_{t^{(k)}_n}$ to $\textbf{w}_{t_{n+1}}$, which in turn guarantees
\begin{equation}
  \lim_{k\rightarrow \infty} \mathbb{P}\big(\textbf{w}_{t^{(k)}_{n}}=(m,n+1,a,b)\big)
= \mathbb{P}\big(\textbf{w}_{t_{n+1}}=(m,n+1,a,b)\big).
\label{convergence}
\end{equation}
Using Eq.~\ref{eqn:definition_conditional_distribution} in 
Eq.~\ref{convergence}, we have

\begin{equation}
\lim_{k\rightarrow \infty} \mathbb{P}\big(\textbf{w}_{t^{(k)}_{n}}=(m,n+1,a,b)\big) 
= \mathbb{P}(m,a,b\,\vert\, t=T_{n+1}).
\end{equation}
The distribution of $\textbf{w}_{t^{(k)}_n}$ is calculated by the
$k^{\rm th}$ iteration in the Algorithm~\ref{alg:update}. We
approximate $\mathbb{P}(m,a,b\,\vert\,t=T_{n+1})$ by the distribution
of $\textbf{w}_{t^{(k)}_n}$ for sufficiently large $k$ and thus
reconstruct $\mathbb{P}(m,a,b,\vert\,t=T_{n+1})$ from
$\mathbb{P}(m,a,b,\vert\,t=T_{n})$.  We iterate this procedure until
$\mathbb{P}(m,a,b,\vert\,t=T_{L})$ is found.  The distribution
$\rho(\Delta T)$ is then found by multiple convolutions of the
exponential distributions with rates $p_{m},\ldots, p_{L}$, weighted
by the probabilities $\mathbb{P}(m,a,b\,\vert\, t=T_L)$ over each $m, a,
b$:
\begin{equation}
  \rho(\Delta T) = \sum_{m} \mathbb{P}\big(m(T_{\rm RNAP})\big)
  \left[\circledast_{j=m}^{L} e^{-p_{j}t} \right](\Delta T)
  \label{eqn:delay-distribution}
\end{equation}
where $\circledast_{j=m}^{L}f_j$ represents sequential convolutions of
functions $\{f_j\}_{j=m}^{L}$; here, $f_{j}(t) = e^{-p_{j}t}$.  An
implementation of the above algorithm in \textsf{Julia} is
available at \url{https://github.com/hsianktin/ttc}.

\subsection{Large system, steady-state approximations}
\label{MFT}

In the limit $L \rightarrow \infty$, we can analyze the system in a
steady-state limit to find a number of useful analytic results.  If we
use the ``center-of-mass'' reference frame, we characterize the system
by the distance $d=n-m$ between the leading ribosome and RNAP. The
dynamics are described by a Markov process on the state space
$(d,a,b)$ described in Fig.~\ref{fig:distance_state_space}. In these
variables, the continuous-time Markov chain admits an equilibrium
distribution $\pi$ and is assumed to be ergodic in the sense that the
fraction of time the system spends at a certain state $A$ is
asymptotically equal to $\pi(A)$. This ergodicity allows us to find
the effective velocity $\overline{V}$ and the fraction of protected
time $F_T$.

With each state $(d,a,b)$ we can associate instantaneous ribosome and
RNAP speeds $V_{\rm rib}$ and $V_{\rm RNAP}$ by the rates of
decreasing and increasing $d$ by one codon, respectively.  For
example, $V_{\rm RNAP}(d=\ell,1,0)=V_{\rm RNAP}(d,a,1)=0$.  Since
ergodicity allows us to find the fraction of time the system is in
state $A$ by its equilibrium probability $\pi(A)$, the effective
velocity can be found by weighting $V_{i}(A), i\in \left\{ \rm RNAP,
rib \right\}$ weighted by $\pi(A)$.  Therefore, at equilibrium, the
effective velocity coincides with the corresponding expected velocity.

%
%

\begin{figure*}[!htb]%
  \centering
  \includegraphics[width=6.4in]{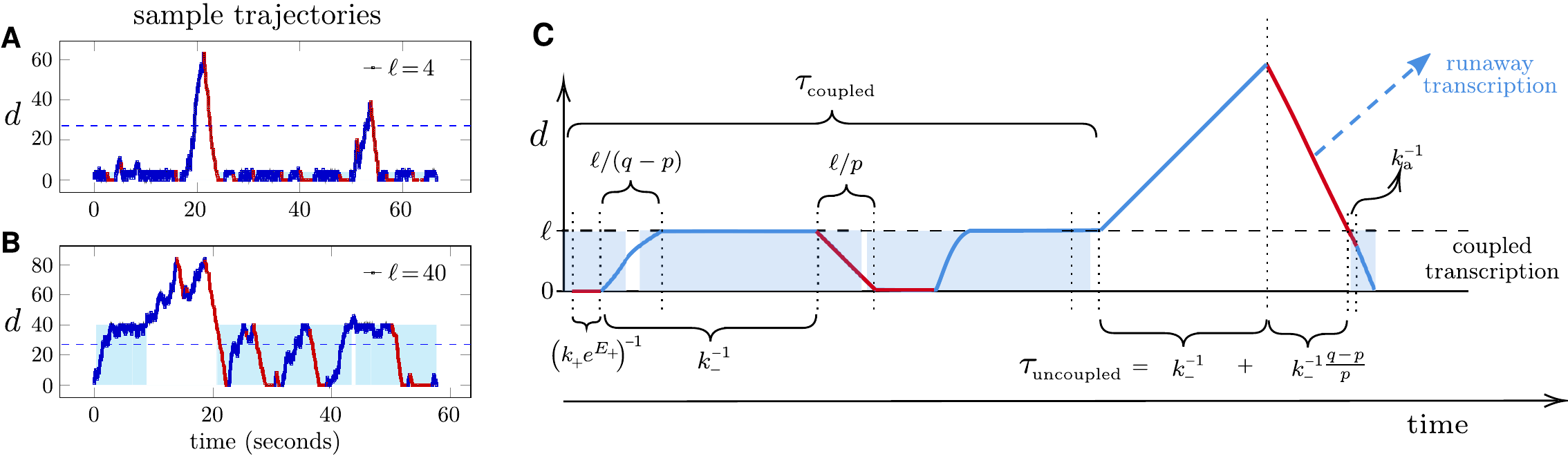}
  \caption{(A) A sample trajectory of distance $d$ between ribosome
    and RNAP as a function of time $t$, with $\ell=4$, $L=1000$, and
    all other parameters equivalent to those used in to generate
    Fig.~\ref{fig:coupling-indices}. (B) A sample trajectory when the
    interaction distance is increased to $\ell = 40$, relevant for
    example, if an intermediate protein such as NusG is involved in
    forming the ribosome-RNAP expressome complex. In (A) and (B),
    blue(red) segments indicate processing(stalled) RNAP. The
    light-blue shaded regions indicate bound ribosome-RNAP
    complexes. (C) Schematic description of distances and estimates of
    transition rates. These estimates assume that microstates within a
    macrostate has reached a local thermodynamic equilibrium in
    $(d,a,b)$-space.}
  \label{fig:sample_low_l_run_away}
\end{figure*}

A sample trajectory of $d$ as a function of time is shown in Fig
\ref{fig:sample_low_l_run_away}A and B. When the ribosome and RNAP are
close ($d\leq \ell$), they can transiently bind and unbind, with dwell
times in each state controlled by $k_{\rm a,d}$.  When the RNAP is
processive, the ribosome lags behind. If the RNAP pauses for a
sufficient time, the ribosome catches up and $d\simeq 0$. Under our
specific set of parameters (large $E_{\rm a}$ and $\bar{q}< p< q$),
$d=\ell$ over most of the trajectory.  Recall that the coupling
constraint prevents the distance to be larger than $\ell$ when
ribosome is bound to RNAP. If $d> \ell$, ribosome and RNAP proceed
independently.

\begin{figure*}[!tbh]%
  \centering
 \includegraphics[width=6.4in]{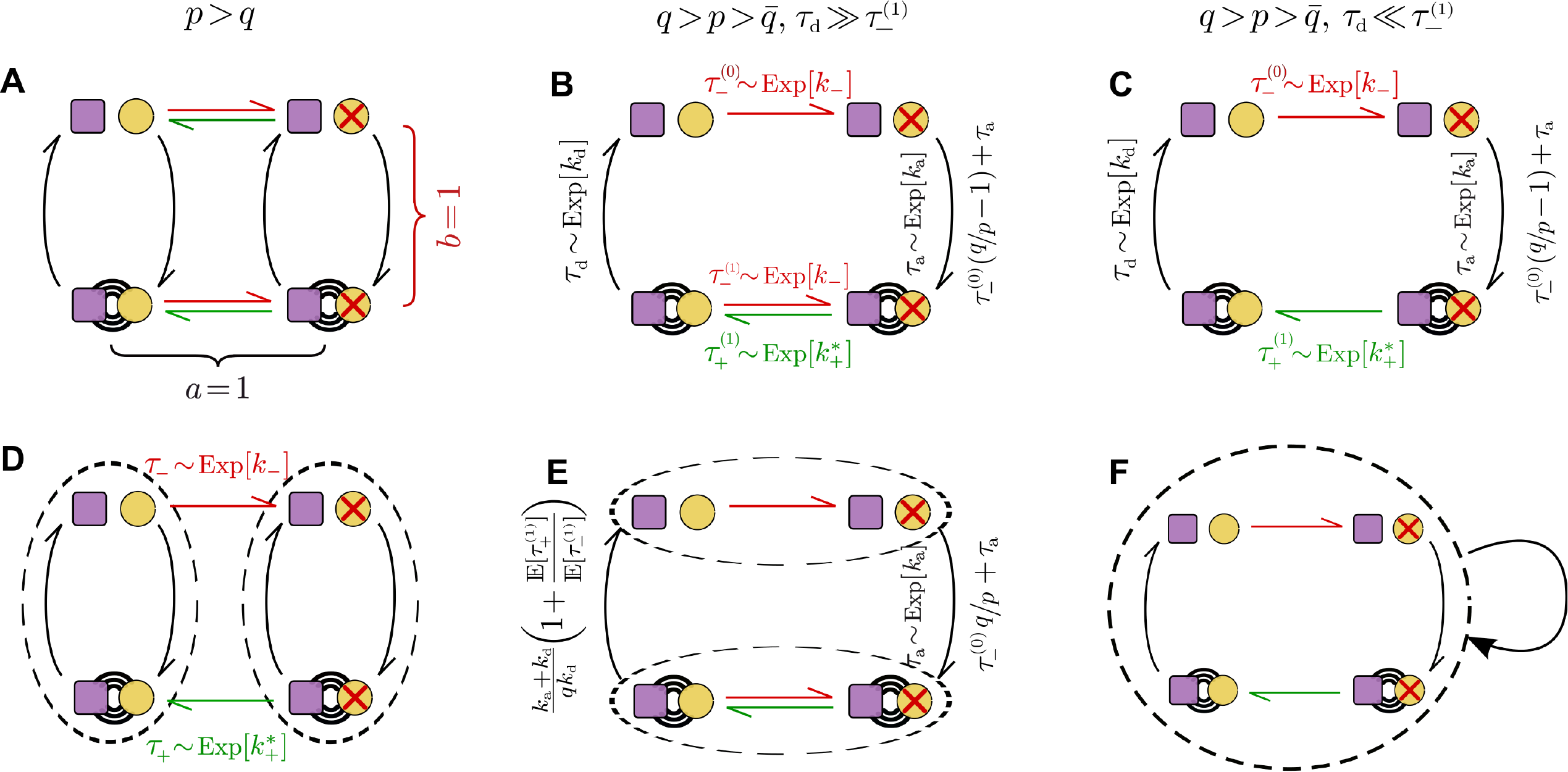}
  \caption{Coarse-grained states of the expressome. (A) A scenario in
    which binding and unbinding are relatively fast and that $p> q$ so
    that the ribosome is fast enough to fully allow binding. This case
    is rarely realized. (B) The limit in which binding is fast, but
    $\bar{q}<p < q$. In this case, when the ribosome and RNAP are
    unbound $(a=0)$, the ribosome falls out of the interaction range
    $\ell$ and cannot immediately rebind. The distance between the
    ribosome and the RNAP remains larger than $\ell$ until the RNAP
    pauses again and ribosome catches up.  When the two machines are
    close, they effectively remain bound. Intermittent unbinding while
    in the processive state quickly leads to separation preventing
    binding until possibly after the next stall event.  (C) Slow
    binding, fast unbinding regime that leads to only transient
    coupling.  The slowdown of transcription is limited in this case.
    (D) The coarse-grained cyclic structure corresponding to the limit
    depicted in (A).  Since binding and unbinding are fast, we can
    lump the ``$a$ states'' together as a single quasi-bound state
    reflecting an equilibrium weighting between bound and unbound. The
    effective rate constant out of this quasistate can be found by
    appropriate weighting of the two component rates.  (E) represents
    the coarse-grained cyclic structure corresponding to the limit
    shown in (B).  Due to strong binding, we assume that the
    transition from bound, $d\leq \ell$ states to unbound, $d>\ell$
    states is slow relative to RNAP stalling/unstalling rates within
    bound states.  Therefore, we group all the $a=1, d\leq \ell$
    states together, and all the $a=0, \d >\ell$ states together.  (F)
    Coarse-grained cyclic structure associated with the limit in
    (C). Scenario (C) is itself a strongly unidirectional cycle that
    we repeat to extract underlying distance traveled. The
    distributions of effective waiting times are indicated, \eg,
    $\tau_{+}^{(a)} \sim \textrm{Exp}[k_{+}^{*}]$ represents an
    exponentially distributed (with rate $k_{+}^{*}$) waiting time in
    the \textit{stalled}, $a$-state before transitioning into a
    processing $a$-state. These rates are derived with an additional
    assumption that $\ell$ is small.}
  \label{fig:cyclic_structures}
\end{figure*}

Fig.~\ref{fig:sample_low_l_run_away}A-B motivates us to lump all the
different states of the system into four representative groups of
microstates:
\begin{itemize}
\item The paused, separated state $(d> \ell, b=1)$.
\item The processive, separated state $(d> \ell, b=0)$.
\item The paused, proximal state, ($d \leq \ell, b = 1$ and $a = 1$ most
of the time).
\item The processive, proximal state, ($d \leq \ell, b = 0$ and $a = 1$
most of the time).
\end{itemize}
In the following, we call these four lumped states as ``macrostates,''
and the states within each macrostate as ``microstates.'' To derive
the effective transition rates between macrostates, we assume that the
microstates within each macrostate reach equilibrium much faster than
the transitions between macrostates. The results are summarized in
Figs.~\ref{fig:sample_low_l_run_away}C and
\ref{fig:cyclic_structures}.


\paragraph*{Traffic jam in associated, processive states.}
As an example of a calculation of transition rates and velocities,
consider the details of the expected velocity of a coupled, processive
expressome. In this particular case, the ribosome and RNAP
intermittently touch ($d=0$) each other. Therefore, they should have
the same effective velocity $\bar{V}$. Suppose that the bound ribosome
is slower than the processing RNAP, $p<q$.  The average speed of the
bound RNAP is thus limited by the speed of ribosome. However, the
ribosome translocates at speed less than $p$ since it is occasionally
blocked by the RNAP.  The equilibrium probability that RNAP and
ribosome are in contact ($d=0$) is given by
\begin{equation}
  \pi(d=0\,\vert\,a=1,b=0) = \frac{1}{\sum_{k=0}^{\ell} (q/p)^{k}}.
\end{equation}
By finding the complementary probability that $0< d\leq \ell$, for
which the ribosome can move forward with rate $p$, we find the
expected expressome velocity
\begin{equation}
 \mathbb{E}\left[{V}\,\vert\,a=1,b=0\right]=
 q\left[\frac{(q/p)^{\ell}-1}{(q/p)^{\ell+1}-1}\right].
\label{eqn:traffic_jam}
\end{equation}
Since $q\sim 30$ codons/s and $p\leq 15$ codons/s, $q/p\sim 2$ and the
relative slowdown is sensitive to $\ell$ with small $\ell$ resulting
in a significant slowdown of the expressome.

\paragraph*{Classification of different scenarios.}
At equilibrium, if the average independent RNAP velocity $\bar{q}$ is
smaller than $p$, the two machines will maintain a significant
probability of proximity and coupling.  However, if $p < \bar q$, the
equilibrium ribosome-RNAP distance $d\to \infty$ and any interaction
will vanish. Thus, we need only consider $p>\bar q$ and discuss the
following scenarios:

\begin{enumerate}
\item The instantaneous speeds satisfy $p\geq q$. Then, the ribosome
  is always within close range of the RNAP and the system freely
  cycles among the four internal macrostates. We may assume that the
  binding and unbinding rates $k_{\rm a}$ and $k_{\rm d}$ are much
  larger than the pausing and unstalling rates $k_{-}$ and $k_{+}$.
\item The instantaneous speeds satisfy $p<q$ and the rate of
  uncoupling $k_{\rm d}$ is slower than the rate of pausing
  $k_{-}$. This system maintains an appreciable probability of being
  coupled. When the RNAP is bound and processive, the distance
  quickly increases until $d\approx \ell$.  Because $k_{-} > k_{\rm
    d}$, the RNAP pauses often before it can break free from the
  ribosome. When the internal states become unbound, the ribosome can
  fall out of the interaction range $\ell$ and cannot immediately
  rebind. This gives rise to an effectively irreversible transition
  from a bound, processive state to an unbound, processive state.
  Rebinding can occur only after the RNAP again pauses, allowing the
  ribosome to catch up. Once this happens, the ribosome and RNAP will
  remain bound for a long time (since $E_{\rm a}$ is large).
\item The instantaneous speeds satisfy $p<q$, but the dissociation
  rate $k_{\rm d}$ is larger than the pausing rate $k_{-}$.  This
  scenario is essentially the same as the previous one, with the only
  difference that the transition from the bound, processive state to
  an unbound processive state is fast and effectively irreversible.
\end{enumerate}

These scenarios can be coarse-grained into different cyclic structures
by grouping states that are connected by reversible reactions, as
shown in Fig \ref{fig:cyclic_structures}D-F. The effective transition
rates are estimated based on the underlying dynamics.
The waiting time distributions can be heuristically estimated as
indicated in Fig.~\ref{fig:sample_low_l_run_away}C and are indicated
in Fig.~\ref{fig:cyclic_structures}.  We can treat the cases depicted
in Figs.~\ref{fig:cyclic_structures}D-F as repeated cycles marked by
when ribosome and RNAP periodically meet each other. Therefore, in the
large-$L$ limit, the effective velocities are approximately equal:
$\overline{V}_{\rm rib} = \overline{V}_{\rm RNAP} \equiv
\overline{V}$.


We analyze the mean and variance of the effective velocity by
estimating the common random time $T$ to complete transcription and
translation of all $L$ codons. First, define $\tau$ as the random time
to traverse one internal-state cycle and $\overline{L}$ as the mean
length traveled in one cycle. To complete a transcript of length $L$,
$\sim L/\overline{L}$ cycles need to be completed. Each cycle can be
considered independent and identically distributed.  The total
variance ${\rm Var}[T]$ and standard deviation $\sigma[T]$ of
completion times is then given by

\begin{equation}
{\rm Var}[T] \equiv (\sigma[T])^{2} \simeq {\rm
  Var}[\tau]\frac{L}{\overline{L}}
\label{eq:iid_variance}
\end{equation}
and the effective velocities are
\begin{equation}
\overline{V}\pm \sigma[V]  \sim \frac{L}{\mathbb{E}[T]\mp \sigma[T]} 
= \frac{\overline{L}}{\mathbb{E}[\tau] \mp \frac{\sigma[\tau]}{\sqrt{L / \overline{L}}}}.
\label{eq:velocity_indeterminancy}
\end{equation}
Thus, it is sufficient to characterize $\overline{L}$ and ${\rm
  Var}[\tau]$ to estimate $\overline{V}$ and its variation in each of
the limits pictured in Figs.~\ref{fig:cyclic_structures}D-F

For the $p>q$ limit (Fig.~\ref{fig:cyclic_structures}D), the overall
velocity is dictated by the velocity of the RNAP and the system can be
approximated by a two-state model in which
\begin{equation}
\begin{cases}
\displaystyle \overline{L} = q \mathbb{E}[\tau_{-}]=\frac{q}{k_{-}}\\
\displaystyle \mathbb{E}[\tau] = \mathbb{E} [\tau_{-}] + 
\mathbb{E}[\tau_{+}]= \frac{1}{k_+^*} + \frac{1}{k_-}\\
\displaystyle {\rm Var}[\tau] = {\rm Var}[\tau_{-}] + {\rm Var}[\tau_+] 
=  \frac{1}{{k_+^*}^2} + \frac{1}{k_-^2}.
\end{cases}
\label{eq:dt_case_a}
\end{equation}
In the limit $\bar{q} < p < q$ and $\tau_{\rm d} \gg \tau^{(1)}_{-}$
(Fig.~\ref{fig:cyclic_structures}E), we apply
Eq.~\ref{eqn:traffic_jam} to find
\begin{equation}
 \begin{cases}
\displaystyle \overline{L} = q \frac{1}{k_-}+ 
\frac{q}{k_{\rm d}} \left[\frac{(q/p)^{\ell}-1}{(q/p)^{\ell+1}-1}\right]\\
\displaystyle \mathbb{E}[\tau] = \mathbb{E}[\tau_{\rm a}] + 
\mathbb{E}[\tau^{(0)}_-] \frac{q}{p}+ 
\mathbb{E}[\tau_{\rm d}^{(1)}] \frac{\mathbb{E}[\tau^{(1)}_+] 
+ \mathbb{E}[\tau^{(1)}_-]}{\mathbb{E}[\tau^{(1)}_{-}]}\\
\displaystyle {\rm Var}[\tau] = {\rm Var}[\tau_{\rm a}] +{\rm Var}[\tau^{(0)}_{-}] 
\frac{q^2}{p^2}+ {\rm Var}[\tau_{\rm d}^{(1)}]
\left(\frac{\mathbb{E}[\tau^{(1)}_{+}] + 
\mathbb{E}[\tau^{(1)}_{-}]}{\mathbb{E} [\tau^{(1)}_{-}]}\right)^2
\end{cases}
\label{eq:dt_case_b}
\end{equation}
In the $q>p>\bar{q}$ and $\tau_{\rm d} \ll \tau^{(1)}_{-}$ limit
(Fig.~\ref{fig:cyclic_structures}F), we also
assume $q \gg k_{\rm a}, k_{\rm d}$ to find

\begin{equation}
\begin{cases}
\displaystyle \overline{L} = q \frac{1}{k_-}+ \frac{q}{k_{\rm d}} 
\left[\frac{(q/p)^{\ell}-1}{(q/p)^{\ell+1}-1}\right]\\
\displaystyle \mathbb{E} [\tau] = \mathbb{E}[\tau_{\rm a}] 
+ \mathbb{E}[\tau^{(0)}_{-}] \frac{q}{p}
+ \mathbb{E}[\tau_{\rm d}^{(1)}] + \mathbb{E}[\tau^{(1)}_{+}] \\
\displaystyle {\rm Var}[\tau] = {\rm Var}[\tau_{\rm a}] + {\rm Var}[\tau^{(0)}_{-}] 
\frac{q^2}{p^2}+ {\rm Var}[\tau_{\rm d}^{(1)}]+ {\rm Var}[\tau^{(1)}_{+}]
\end{cases}
\label{eq:dt_case_c}
\end{equation}
\paragraph*{Estimation of the effective velocity and fraction of protected time.}
It turns out that with realistic parameter values, our metrics are
rather insensitive to the magnitudes of $k_{\rm a}$ and $k_{\rm
  d}$. Therefore, we focus on the cases $\tau_{\rm d} \gg
\tau_{-}^{(1)}$ and $q \gg k_{\rm a}$ or $k_{\rm d}$.  In the case
$k_{\rm d}\ll k_{-}$ and $k_{\rm d} \ll k_{\rm a}$, the ribosome and
RNAP are in molecular contact most of the time.  For the sake of
simplicity, we consider the extreme limit $E_{\rm a} \rightarrow
\infty$.  and simply evaluate the effective velocity for the coupled
expressome. The mean/effective velocity $\overline{V}$ can be
constructed using Eq.~\ref{eq:dt_case_b} and simplified to
\begin{equation}
  \overline{V} \approx q \left[\frac{(q/p)^{\ell}-1}{(q/p)^{\ell+1}-1}\right] 
\frac{k_{+}^{*}}{k_{+}^{*}+k_{-}}.
  \label{eq:velocity_case_b}
\end{equation}
For sufficiently large $q$ and $\ell$, the coefficient
$\left[\frac{(q/p)^{\ell}-1}{(q/p)^{\ell+1}-1}\right] \sim p/q $ and
$\overline{V} \approx pk_{+}^{*}/(k_{+}^{*}+k_{-})$ as expected.
Since $k_{+}^{*} \geq k_{+}$, the effective velocity has a lower bound
\begin{equation}
 \overline{V}\geq \frac{pk_{+}}{k_{+}+k_{-}}\geq \frac{\bar{q} k_{+}}{k_{+}+k_{-}}.
  \label{eq:velocity_case_b-m}
\end{equation}

\paragraph*{Estimation of the protected fraction.} Under the same 
assumption that $\bar{q} < p < q$, $\tau_{\rm d} \gg \tau_{-}^{(1)}$,
it is possible to estimate the protected fraction by investigating the
equilibrium distribution of distances $d$.  Here, we need to
separately discuss two scenarios as shown in
Figs.~\ref{fig:sample_low_l_run_away}A and B, respectively. The two
different cases are characterized by the length of interaction $\ell$.

In the first case, $\ell/(q-p) \ll k_{-}^{-1}$. Consequently, in the
bound, processive state, the ribosome-RNAP distance $d\approx \ell$
for the most of the time. 
%
%
In the second case, $\ell/(q-p) \gg k_{-}^{-1}$ and $\ell > \ell_{\rm
  p}$. Consequently, in the bound, processive state, the ribosome
spends most of its time separated $0<d<\ell$. Only occasionally,
$d=\ell$ before RNAP unpauses again and RNAP is able to break away
from the lagging ribosome.

If the first case holds, unprotected states arise only when the system
is unbound and $d>\ell$. Here, we consider a simple asymmetric random
walk starting from $d=\ell$ that increases to $d+1$ with rate $q$ and
decreases to $d-1$ with rate $p$.  There are three stopping times that
come into play. $\tau_{-1}$ is the first time $d_t=\ell-1$.
$\tau_{\ell_{\rm p}}$ is the first time $d_{t}=\ell_{\rm p}$, and
$\tau_{-}$ is the time to first RNAP pausing.

We heuristically estimate the probability that $\tau_{\ell_{\rm p}} <
\tau_{-1}$, $\tau_{-}$ as well as the mean duration of exposure
conditional on $\tau_{\ell_{\rm p}} < \tau_{-1}, \tau_{-}$. The
average of the largest distance $d_{\rm max}$ during the whole process
is given by $(q-p)/k_{-}$. The standard deviation of $d_{\rm max}$ is
determined by the square root of the average number of steps
$\sqrt{(p+q)/k_{-}}$. Then, roughly $d_{\rm max}$ follows a normal
distribution with mean $(q-p)/k_{-}$ and standard deviation
$\sqrt{(p+q)/k_{-}}$. The probability that there is an unprotected
duration is given by $\mathbb{P}(d_{\rm max}> \ell_{\rm p} -
\ell)$. Due to the memoryless property of the exponential
distribution, the average duration conditional on $d_{\rm max} >
\ell_{\rm p} - \ell$ is essentially the same as the average duration
of the whole uncoupled event $\frac{q}{p k_{-}}$.  Therefore, and
estimate of the mean protected time fraction is
\begin{equation}
  \mathbb{E}[F_T]  \sim 1 -  
\frac{\frac{q}{p k_{-}}\mathbb{P}(d_{\rm max} > \ell_{\rm p} - \ell)}
{\frac{q}{p k_{-}} 
+ \frac{k_{\rm a} + k_{\rm d}}{qk_{\rm d}} \frac{k_{+}^* + k_{-}}{k_{+}^*}}
\approx  1 -  \frac{\frac{q}{2pk_{-}}
\left[1-\textrm{erf}\bigg(\frac{\ell_{\rm p}- 
\ell- (q-p)/k_{-}}{\sqrt{2(p+q)/k_{-}}}\bigg)\right]}
{\frac{q}{p k_{-}} + \frac{k_{\rm a} + k_{\rm d}}{qk_{\rm d}} 
\frac{k_{+}^* + k_{-}}{k_{+}^*}}
 \label{eq:f_t_small_l}
\end{equation}

For the second case, $\ell/(q-p) \gg k_{-}^{-1}$ and \added{$\ell >
  \ell_{\rm p}$,} we are primarily interested in the bound states
since as $\ell$ increases, the chances that unbound states arise
decrease. However, since $\ell > \ell_{\rm p}$, even in the bound
state, there is a chance that $d > \ell_{\rm p}$.  Estimation of the
probability that the exposed state is visited follows a similar
argument as the previous calculation where we examined the
distribution of $d_{\rm max}$. The main difference is that the
asymptotic distribution of $d_{\rm max}$ is now different from the
normal distribution due to $\ell > \ell_{\rm p}$. This also
changes the duration of exposed states. For simplicity, we 
consider only the $\ell \to \infty$ limit to find
\begin{equation}
\mathbb{E}[F_T]  \sim 1 -
\frac{qk_{+}^{*}}{2(qk_{+}^{*}+pk_{-})}\left[1-\textrm{erf}\bigg(
  \frac{\ell_{\rm p} - (q-p)/k_{-}}{\sqrt{2(p+q)/k_{-}}}\bigg)\right].
\label{eq:f_t_large_l}
\end{equation}
%
%

Our preliminary analyses suggest that in the short interaction range
limit, the quantity $\frac{\ell_{\rm p} - \ell -
  (q-p)/k_-}{\sqrt{2(p+q)/k_{-}}}$ plays a significant role in
determining $F_T$, while in the long interaction range limit,
$\frac{\ell_{\rm p} - (q-p)/k_-}{\sqrt{2(p+q)/k_{-}}}$ plays a similar
role.



\subsection{Variability of the protected-time fraction $F_{T}$}
\label{variations}

Fig.~\ref{fig:coupling-indices}E plotted only the expected protected
fraction. Since $F_{T}$ we generated via the full stochastic
simulation, the variability of $F_{T}$ is also of interest. Here, we
plot the standard deviation $\sigma[F_{T}]$ versus simulated values of
$F_{T}$ to show that it agrees qualitatively well with
$\sqrt{\mathbb{E}[F_T]\big(1-\mathbb{E}[F_T]\big)}$.


\begin{figure}[!tbh]%
\centering
\includegraphics[width=3in]{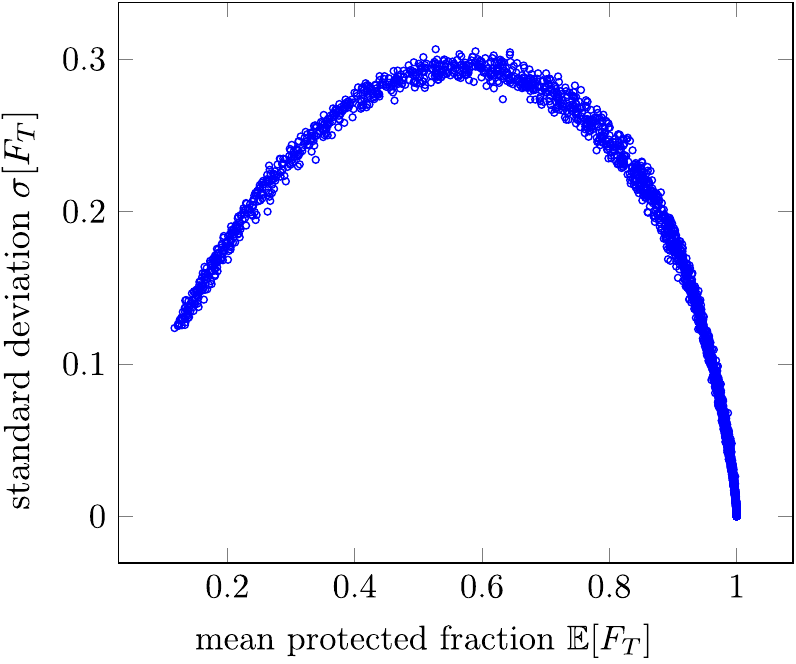}
\caption{Standard deviation $\sigma[F_T]$ as a function of
  $\mathbb{E}[F_T]$ computed using different values of $(p,q)$. All
  other parameter values are those used in
  Fig.~\ref{fig:coupling-indices}E.}
\label{fig:mean_std_F_T}
\end{figure}

\subsection{Effects of interaction length $\ell$}
\label{interaction_ell}
The interaction length $\ell$ is one factor that influences
coupling-induced slowdown, as indicated by
Eq.~\ref{eq:velocity_case_b}.  The interaction length is not a
significant contributing factor to slowdown because the factor
$\left[\frac{(q/p)^{\ell}-1}{(q/p)^{\ell+1}-1}\right]$ is already
$\sim 1$ when $\ell \sim 5$. This factor is small only when
$\ell\approx 0$. Since $\ell$ takes on integer values this slowdown
factor never really really becomes very small.  On the other hand, the
interaction length $\ell$ also dictates the distribution of $d$
conditioned on $a=1$. For example, if $\ell/p \gg k_{-}$, then the
most probable distance between ribosome and RNAP will be $d = \ell$.

\begin{figure}[!tbh]%
  \centering
  \includegraphics[width=4.8in]{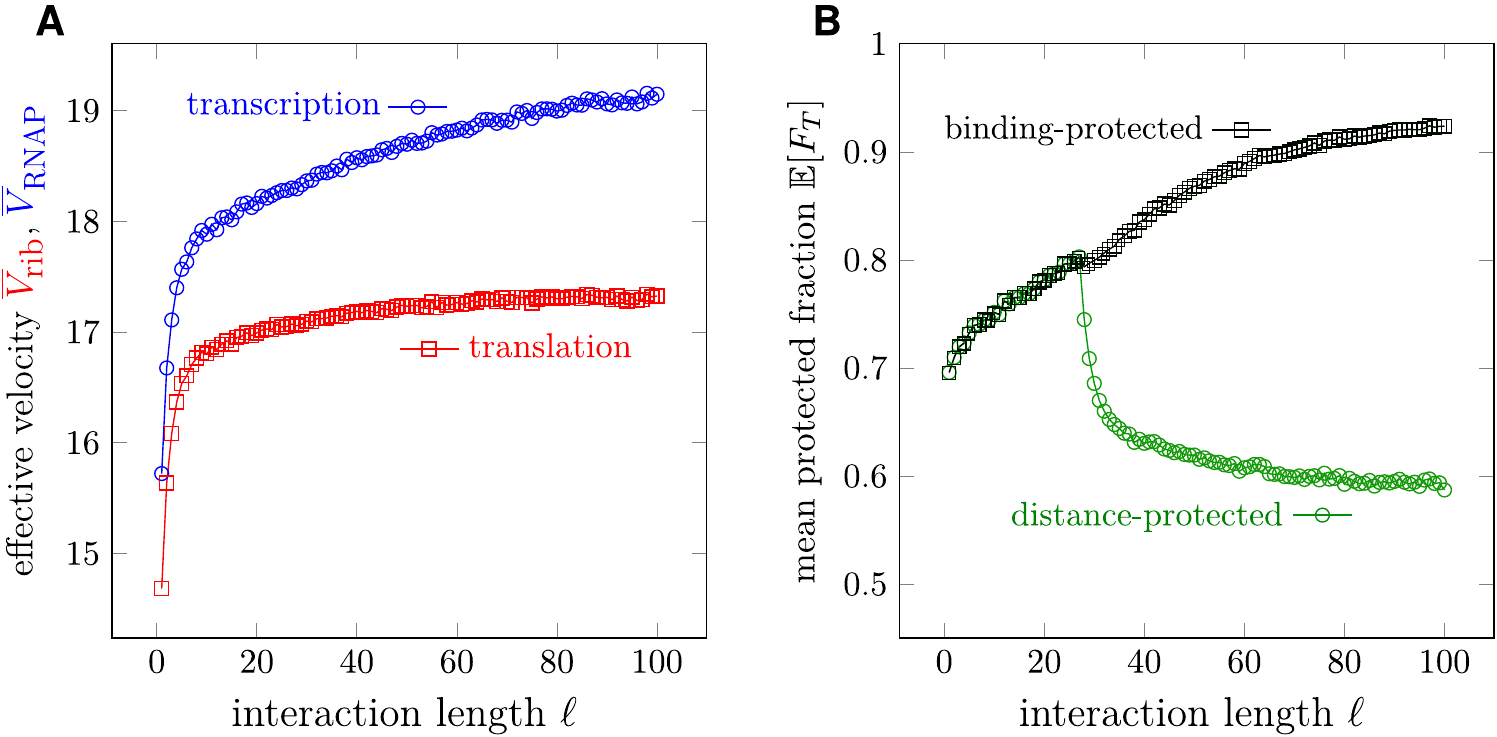}
  \caption{Functional consequences of varying interaction length
    $\ell$. Unspecified parameters are the same as those used in
    Fig.~\ref{fig:coupling-indices}. (A) Effective velocities as a
    function of $\ell$. Note the sharp drop when $\ell \rightarrow 0$.
    (B) Bifurcation of $\mathbb{E}[F_T]$ as a function of $\ell$ due
    to different definitions of protection. Other parameters are the
    same as those used in the Fig.~\ref{fig:coupling-indices}. When
    $\ell < \ell_{\rm p}=27$, the two definitions of $F_T$ agree. When
    $\ell > \ell_{\rm p}$, the $\mathbb{E}[F_T]$ based purely distance
    is drastically lower as $\ell$ increases.}
   \label{fig:supp_3}
\end{figure}

We have found an interesting ``bifurcation'' in effective velocity and
mean protected times when the interaction distance $\ell > \ell_{\rm
  p} = 27$, the mRNA footprint length of a transcription terminator
such as Rho. If $\ell < \ell_{\rm p}$, protection by the ribosome can
be thought of as being purely due to steric exclusion effects; once $d
> \ell_{\rm p}$, protection is lost. However, if $\ell>\ell_{\rm p}$
one can consider a ``binding-based'' protection that requires either
$d\leq \ell_{p}$ \textit{or} $a=1$ for protection.  In this case, even
if $\ell > d>\ell_{\rm p}$, there can be protection due to
binding-mediated conformational shielding of the intervening mRNA that
makes it inaccessible to termination factors.  The different criteria
for protection lead to drastically different levels of protection
provided by the ribosome, as shown in Fig.~\ref{fig:supp_3}.

\end{document}